\documentclass[review]{elsarticle}

\usepackage{amsmath}
\usepackage{amssymb}
\usepackage{algorithm}
\usepackage{comment}
\usepackage{natbib}
\usepackage{mathtools}
\usepackage{graphicx}
\usepackage{subcaption}
\usepackage{algorithmic}
\usepackage{varwidth}
\usepackage{xcolor}
\usepackage{booktabs}
\usepackage{pgfplotstable}

\usepackage{lineno,hyperref}
\modulolinenumbers[5]
\definecolor{RevCol}{RGB}{140, 20, 160}
\journal{peer review}

\makeatletter
\def\ps@pprintTitle{%
	\let\@oddhead\@empty
	\let\@evenhead\@empty
	\def\@oddfoot{}%
	\let\@evenfoot\@oddfoot}
\makeatother

\begin{document}

\begin{frontmatter}

\title{Probabilistic electric load forecasting through \\ Bayesian Mixture Density Networks}

\author[mymainaddress,mysecondaryaddress]{Alessandro Brusaferri\corref{mycorrespondingauthor}}
\ead{alessandro.brusaferri@stiima.cnr.it}
\author[mysecondaryaddress]{Matteo Matteucci}
\author[mymainaddress]{Stefano Spinelli}
\author[mymainaddress]{Andrea Vitali}
\cortext[mycorrespondingauthor]{Corresponding author}
\address[mymainaddress]{CNR, Institute of Intelligent Industrial Technologies and Systems for Advanced Manufacturing, via A. Corti 12, Milan, Italy}
\address[mysecondaryaddress]{Politecnico di Milano, Department of Electronics, Informatics and Bioengineering, via Ponzio 34/5, Milan, Italy}

\begin{abstract}
Probabilistic load forecasting (PLF) is a key component in the extended tool-chain required for efficient management of smart energy grids. Neural networks are widely considered to achieve improved prediction performances, supporting highly flexible mappings of complex relationships between the target and the conditioning variables set. However, obtaining comprehensive predictive uncertainties from such black-box models is still a challenging and unsolved problem. In this work, we propose a novel PLF approach, framed on Bayesian Mixture Density Networks. Both aleatoric and epistemic uncertainty sources are encompassed within the model predictions, inferring general conditional densities, depending on the input features, within an end-to-end training framework. To achieve reliable and computationally scalable estimators of the posterior distributions, both Mean Field variational inference and deep ensembles are integrated.  Experiments have been performed on household short-term load forecasting tasks, showing the capability of the proposed method to achieve robust performances in different operating conditions.
\end{abstract}

\end{frontmatter}


\section{Introduction}
Electric load forecasting (LF) is an essential tool for the optimal operation and planning of energy grids. In particular, the short-term load forecasting (STLF) - i.e., the prediction from several minutes up to one week ahead - is crucial for overall system reliability, to maintain stable balance between supply and demand, and to support effective dispatching and commitment of generation units \cite{HE2020114396}. In financial terms, it has been estimated that a 1\% increase of the load prediction error cause losses to utilities of hundreds thousand dollars per GW peak \cite{hong2016probabilistic}.

Therefore,  a broad set of load forecasting methods have been investigated over the years, often clustered in two major families, namely statistical and artificial intelligence based techniques. A non exhaustive list includes linear auto-regressive models and related extensions (e.g., ARMAX, GARCH, etc.) \cite{weronbook}, exponential smoothing \cite{expoSmoth}, generalized additive models \cite{GAILLARD20161038}, Gaussian Process \cite{YANG2018499}, gradient boosting \cite{BENTAIEB2014382}, support vector machines \cite{1350819}, random forest \cite{LAHOUAR20151040}, fuzzy logic \cite{5685606}, neural networks \cite{8372953} and hybrid models \cite{7124541}. More detailed reviews and comparison of these LF approaches can be found in \cite{8322199},\cite{HAHN2009902},\cite{hong2016probabilistic}. Modern neural network (NN) architectures and deep learning (DL) techniques are being widely considered nowadays for LF, exploiting their enhanced representation capabilities and the increasing availability of tools and highly powered computational resources, leading them amendable also for big data applications\cite{8372953}.

In the recent years, the electric load volatility has increased rapidly and unprecedented challenges have been introduced by the augmented penetration of renewable energy sources, the adoption of extended demand response programs and liberalized markets with increasingly complex pricing policies \cite{BRUSAFERRI20191158}. In such a context, a reliable characterization of the uncertainties associated with the prediction is fundamental to achieve effective decision making processes including detailed risk assessments \cite{8988175}, stochastic optimizations \cite{MUNKHAMMAR2021116180}, optimal production scheduling \cite{RAMIN2018622}, etc. Therefore, an increasing research effort is being dedicated to the development of probabilistic extensions to conventional LF techniques, i.e. probabilistic load forecast (PLF) (see e.g., \citealp{hong2016probabilistic} for a detailed review).

Despite the significant results reached in terms of point forecasting error reduction, quantifying predictive uncertainty in NNs is a challenging and yet unsolved problem \cite{pmlr-v80-kuleshov18a}. In fact, model parameters are typically estimated in practical regression settings by minimizing sum of squares errors over training data, resulting in predictions of the conditional expectations of the targets in out of samples conditions (i.e, forecasted loads over the prediction horizon).

Prediction uncertainity assessment is particularly critical for neural networks. Indeed, while calibrated estimates are mandatory for the safe exploitation of forecasting models in real-world applications, recent studies have demonstrated that conventional deep learning approaches are prone to overconfident (i.e, mis-calibrated) outcomes \cite{guo}.  Basically, deployed models do not convey proper indications regarding “when they should/should not be trusted” sample-wise, due e.g., to the relative distance of the test data instance from the training regions within the overall feature space \cite{NEURIPS2019_8558cb40}.

To accomplish this goal, both aleatoric and epistemic sources of uncertainties have to be properly addressed within the PLF framework \cite{gal1}. The former captures the stochasticity inherent in the observations, resulting in prediction errors which cannot be reduced regardless of the quantity of available data and model quality. The latter accounts for the model uncertaininty, expected to be explained away by obtaining more representative data. Indeed, epistemic uncertainty is particularly relevant when tackling empirical datasets of limited size, including sparse samples \cite{wilson}.

A broad spectrum of approaches have been developed to asses the aleatoric uncertainty in neural network models for PLF, including ex-post analysis techniques assuming input-independent covariances, prediction intervals quantification, Lower Upper Bound Estimation and Quantile Regression (QR) based methods (see \cite{hong2016probabilistic} for a detailed review of the major methods in this field up to 2016). Various extensions to Quantile Regression Neural Networks (QRNN) have been investigated over the past five years. For instance, a Pinball loss function is exploited in \cite{wang2019probabilistic} and \cite{8810811} to guide learning and extract quantiles from recurrent Long Short Term Memory and Convolutional networks respectively. Authors in \cite{9024710}  introduced embedding layers in QRNNs to address categorical features. More computational amendable procedures are investigated in  \cite{zhang2018improved} to mitigate the lack of scalability of conventional QRNN to datasets of reasonable size. A  Least Absolute Shrinkage and Selection Operator (LASSO) based quantile forecast combination strategy is proposed in \cite{yang2019deep}, following a multitask learning approach.

Beyond prediction intervals and quantiles expression, a full statistical characterization of the inherent stochasticity of the electricity load, depending on the input features, can be achieved through conditional distributions \cite{afrasiabi2020deep}. Discrete intervals and summaries (e.g, tendency, dispersion, etc.) can be considered as sub-cases. Therefore, an increasing research effort is being devoted to techniques to transform the outputs give by previous QRNN into probability density forecasts \cite{HE2020114396}. Both non-parametric and parametric methods has been investigated for such purpose. Specifically, authors in \cite{guo2018deep} explored a Kernel Density Estimation (KDE) technique with Gaussian kernel functions to post-process feed-forward QRNNs predictions. A multi-step algorithm is proposed in  \cite{he2019electricity}, exploiting LASSO-QRNNs training to include input features selection, followed by a KDE stage based on Epanechnikov and Gaussian kernels. 
A perturbation search method is investigated in \cite{zhang2020load} aimed to combine multiple KDE transformed QR components, weighted by the Continuous Ranked Probability Score (CRPS). A parametric Gaussian approximation is proposed in \cite{li2019combining} to convert the QR models outputs into probability densities, thus avoiding the high sensitivity issue of KDE to bandwidth hyperparameter tuning, especially in lower samples regimes around the forecast distribution tails. Moreover, QRNNs are combined with QR-Gradient Boosting and Gaussian Process Regression to minimize the overall CRPS - by solving a linearly constrained quadratic programming problem -, obtaining superior PLF performance with reference to previous QR techniques. In fact, despite the simplifying Gaussian assumption at single model level, the overall framework results in a mixture distribution capable to approximate general densities, as required to address complex stochastic patters in load data. 
A PLF approach aimed to directly infer full conditional densities have been recently introduced in \cite{afrasiabi2020deep}, thus avoiding the need to pass through QRNNs transformations. To this end, a Mixture Density Network (MDN) have been adopted. Constituting more a class of techniques for flexible density estimation than a specific NN architecture, MDNs can approximate complex conditional distributions - including e.g., skewed and multi-modal patterns -  up to arbitrary accuracy \cite{Bishop:2006:PRM:1162264}. Experiments have been performed on residential households consumption, reporting improved performances with reference to state of the art methods, including the combination technique proposed in \cite{li2019combining}. It is worth noting that, indeed, authors in \cite{li2019combining} envisioned extensions at single model levels overriding the restricting Gaussian assumption. Besides, conventional MDN inference can suffer computational problems (i.e., mode collapse) and poor generalization, requiring proper extensions at both architectural and learning machinery levels \cite{Bishop:2006:PRM:1162264}, \cite{afrasiabi2020deep}.  

Unlike the substantial amount of research devoted to aleatoric uncertainity characterization within NN-PLF, the epistemic counterpart is still largely unexplored. 
Notably, the integration of the epistemic uncertaininty contribution within NN predictions can be thoroughly addressed under the Bayesian framework \cite{liberty}. 
Specifically, Bayesian Neural Networks (BNN) replace the point estimates in the parameters space (often inferred using maximum likelihood or maximum a posteriori by frequentist approaches) with entire posterior distributions computed using the Bayes rule, thus offering a principled approach to capture epistemic uncertainty as well as an intrinsic regularization effect \cite{Bishop:2006:PRM:1162264}. Indeed, as NNs typically have highly diffuse likelihoods and broad valleys in the loss landscape, different parameter settings produce various predictive functions consistent with the empirical observations \cite{pmlr-v115-izmailov20a}. Then, by following the Bayesian inference approach, output distributions can be obtained through marginalization, thus conveying model confidence from the features space up to the network predictions \cite{wilson}. However, standard inference techniques commonly adopted for simple Bayesian regression models and small data regimes are not computationally feasible for deep learning applications. Therefore, a lot of research have been dedicated to the development of scalable and effective approximation methods (see e.g., \cite{10.1145/3409383} and references therein).

Despite the BNN developments briefly summarized above, which have been mainly deployed within the computer science field, Bayesian deep learning techniques have still attracted minor attention in the electricity load forecasting context. To the best of our knowledge, the only previous works focusing epistemic uncertaininty through BNNs for PLF are \cite{8372953} and \cite{8743433}, but still limited to the simplifying Gaussian assumption for the aleatoric counterpart. We found a single study investigating epistemic uncertainty in MDNs, performed within the autonomous driving research field \cite{8462978}. However, the proposed estimation method is reduced to the Gaussian aleatoric uncertainity sub-case. Summarizing, the exploration of comprehensive predictive uncertainties in MDN-PLF models through Bayesian deep learning extensions is still lacking in the literature. 

\subsection{Contributions and organization of the paper}
Starting from the key research results introduced, and considering the reported open challenges, the main scope of this work is to support the development of probabilistic extensions to conventional neural network based load forecasting techniques, by augmenting their flexible representation capabilities with comprehensive uncertainty characterizations. To this end, we proposed a novel approach to probabilistic load forecasting based on Bayesian Mixture Density Networks. \\
Specifically, major contributions of this paper are the following:
\begin{itemize}
\item an enhanced Bayesian Mixture Density Network formulation is conceived to capture both aleatoric and epistemic uncertainty counterparts within model predictions, while inferring complex conditional distributions.
\item	To achieve reliable function space posteriors, while retaining scalable training procedures,
we integrate Mean Field variational inference and deep ensembles, providing complementary approximation capabilities at both local (i.e, around single-basins) and global (i.e, covering multiple-modes) scale within the Bayesian inference machinery. Besides, a tempered posterior is incorporated in the inference process  to address potential over-regularization of large Bayesian neural networks under limited data settings, balancing model capacity to the effective amount of observations.  
\item	An end-to-end network learning is performed, bypassing ex-post output transformations requirements, so to discover the latent functional relation to conditioning variables, characterize inherent load stochasticity, and convey parameters uncertainity in a single PLF framework.
\item	Experiments are performed over real applications with heterogeneous behaviors, showing improved probabilistic forecast performance with reference to state of art techniques. Specifically, we address STLF tasks at individual household scale, characterized by highly volatile patterns.
\end{itemize}
The ultimate aim is to further foster the development of Bayesian deep learning techniques in the load forecasting context, as underlying mechanisms to convey the uncertainties associated with neural networks predictions, thus supporting reliable decision making processes.\smallskip \\
The rest of the paper is structured as follows. Section 2 starts introducing the load forecasting problem from a general Bayesian inference perspective. Then, each element of the proposed approach is described, including the specification of the parameterized mixture distribution output, the overall network architecture, the developed approximate inference and training techniques, as well as the adopted scores for PLF performance evaluations. Section 3 analyzes the  STLF case studies considered, providing a detailed description of the experimental setups and configurations, and reporting the results achieved. Section 4 summarizes conclusion and the envisioned future extensions.

\section{Methods}
\subsection{Preliminaries: from frequentist to Bayesian neural-network-based LF}
Conditional density estimation targets the identification of reliable representations of the underlying data generating process, for the purpose of making analysis and predictions in test conditions \cite{Bishop:2006:PRM:1162264}. When regression tasks are approached through neural network models, inference is typically performed by maximizing the likelihood of the available observations over parameterized distributions, optionally adding regularization terms to mitigate overfitting \cite{Goodfellow-et-al-2016}.
When a homoskedastic Gaussian form is assumed, this leads to the following optimization problem:
\begin{equation}
\small
    \omega^*= \arg\underset{\omega}{\min}\sum_{n=1}^N -\log p(\mathrm{y}_n|\mathrm{x}_n, \omega)=\arg\underset{\omega}{\min}\sum_{n=1}^N -\log\frac{1}{\sqrt{2\pi\sigma_y}}e^{-\frac{1}{2{\sigma_y}^{2}}(f_{NN}(\mathrm{x}_n, \omega)-\mathrm{y}_{n}))^2}
\end{equation}
which is often referred to as negative log-lokelihood, reducing to the common sum of squares minimization. To lighten notation, we employ a single output form, where the dataset $\mathcal{D}={\left\lbrace \mathrm{x}_n,\mathrm{y}_n\right\rbrace }_{n=1}^N$ comprises $N\in \mathbb{Z}^+$ 
independent and identically distributed (i.i.d.) realizations of the input features and dependent variables pairs in the training set, with $\mathrm{x} \in \mathbb{R}^{n_x}, \mathrm{y} \in \mathbb{R}$, while $w \in \Omega\subseteq \mathbb{R}^{n_{\omega}}$. Hence, the network learns to approximate the conditional mean in the target space given the values of the conditioning variables (i.e., ${f}_{NN}\left({\mathrm{x}}_{n}, {\omega}\right)$), which depends on the parametrization at the local minimizer where the training algorithm converged.\\
Under this setting, the variance parameter $\sigma_y \in \mathbb{R}^+$ is usually estimated through the residual over the validation subset:
\begin{equation}
{\sigma}_y^{2}=\frac{1}{N_v} \sum_{n=1}^{N_v} \left[{f}_{NN}\left({\mathrm{x}}_{n}, {\omega}^{*}\right)-\mathrm{y}_{n}\right]^{2}
\end{equation}
thus providing an average prediction variance. 
Heteroskedastic normal extensions can be obtained by parametrizing the variance parameters through dedicated network outputs. However, while the exploitation of the sum of squares loss does not strictly require a Gaussian form for the underlying conditional distribution, the network is unable to differentiate it from alternatives which do have the same statistics \cite{bishop94}.

Besides the consistent specification of the parameterized distribution, the generalization capabilities of network models are particularly crucial in high-dimensional density estimation settings. Indeed, the learning machinery does not have access to the ground truth conditional distribution, while gathering realizations from exactly the same features values is very unlikely \cite{Makansi_2019_CVPR}.

Regardless of the class of distribution adopted, maximum likelihood (or even extended maximum a posteriori) approaches to neural network training infer point estimates in the parameters space, thus leading to predictive models agnostic to epistemic uncertainty \cite{9150658}. 
A principled approach to encompass epistemic uncertainty with the network is given by Bayesian statistics \cite{Neal:1996:BLN:525544}. Specifically, the weights are intrinsically considered as stochastic variables, represented through an overall posterior distribution $p(\omega|\mathcal{D})$ given by the application of the Bayes theorem:
\begin{equation}
p(\omega|\mathcal{D})=\frac{p(\mathcal{D}|\omega)p(\omega)}{\int_\Omega p(\mathcal{D}|\omega)p(\omega)}
\label{post_dist}
\end{equation}
where the numerator factorizes in the likelihood $p(\mathcal{D}|\omega)$ times the network parameters prior $p(\omega)$, and the denominator constitute the evidence over the available dataset.
Then, network parameters distributions are marginalized into the function space posterior:
\begin{equation}
p(\mathbf{y}|\mathbf{x},\mathcal{D})=\int_\Omega p(\mathbf{y}|\mathbf{x},\omega)p(\omega|\mathcal{D})d\omega
\label{post_int}
\end{equation}
Hence, testing time predictions are performed by accounting for the epistemic uncertainity within a Bayesian Model Average, rather then relying on a single setting of the weights (i.e., one hypothesis) as in conventional training methods, which cannot be optimally chosen given limited amount of data \cite{minka}.
It is worth noting that maximum likelihood approaches to neural networks training can be considered a very basic approximations of the integral, employing uniform priors and Dirac-delta posteriors, thus betting on single hypothesis having densities concentrated in point-masses in the parameters space \cite{10.1145/3409383}.

Given the theoretical background of Bayesian neural networks, the achievement of reliable PLF models requires the definition of proper likelihood/prior forms, considering the specific requirements of probabilistic load forecasting problems at hand. Besides, a computationally amendable inference technique has to be developed to enable the exploitation of BNNs in practical conditions.
Such issues are detailed and tackled within the following subsections.

\subsection{Modeling arbitrary conditional distribution in PLF through MDNs}
Following a Bayesian approach, the first step is the specification of the likelihood function.
To enable the estimation of general conditional distribution shapes, thus characterizing the aleatoric uncertainity in the predictions, we employ the architectural paradigm of Mixture Density Networks. Specifically, the linear output layer of the network -  included in conventional LF neural models assuming a Gaussian distribution - is replaced by a probabilistic layer implementing a mixture model, whose parameters are flexibly mapped by the lower layers in the architecture, depending on the specific values of the conditioning features. 
Starting from the general MDN concept, a broad range of neural PLF models can be designed. A first choice regards kernels characterization and covariance matrices. Various alternatives have been considered for different application contexts in the literature (see e.g. \cite{817982},\cite{Makansi_2019_CVPR},\cite{journals/corr/Graves13},\cite{8988175}). 
In this work, we developed spherical Gaussian kernels, providing a more computationally scalable alternative to the full covariance forms (e.g., using lower triangular components in Cholesky decompositions), while still supporting general conditional densities approximations to arbitrary accuracy \cite{2018arXiv180701987G},\cite{bishop94}.
Formally, the MDN kernels in the output distribution are defined as follows:
\begin{equation}
\phi_{k}(\mathrm{y} | \mathrm{x}, \omega)=\frac{1}{(2 \pi)^{1 / 2} \sigma_{k}(\mathrm{x})} \exp \left\{-\frac{\left\|\mathrm{y}-\mu_{k}(\mathrm{x})\right\|^{2}}{2 \sigma_{k}(\mathrm{x})^{2}}\right\}
\end{equation}
where $\mu_k(\mathrm{x})\in\mathbb{R}$ and $\sigma_k(\mathrm{x})\in\mathbb{R}^+$ constitutes the input conditioned mean and variance parameters of the $n_k$ component in the mixture.
Hence, the overall output density of the LF model results:
\begin{align}
&p(\mathrm{y} | \mathrm{x}, \omega)=\sum_{k=1}^{n_k} \alpha_{k}(\mathrm{x}) \phi_{k}(\mathrm{y} | \mathrm{x}), \quad\text { with: } \sum_{k=1}^{n_k} \alpha_{k}(\mathrm{x})=1 
\end{align}
where $\alpha_k\mathrm{(x)} \in\mathbb{R}$ represents the mixing coefficients, weighting the components in the superposition. Here, we lightened notation by implicitly considering the dependence on the network parameters $\omega$.

To achieve a correct GMM parametrization through the network, the last hidden layer has to be properly configured to guarantee mixing coefficients residing on the $n_k$-dimensional simplex and positive definite variances \cite{Bishop:2006:PRM:1162264}. For the former, we adopt a parameterized categorical distribution, thus constraining the weighing proportions, employing a softmax function during predictions. Regarding the latter, on the output variance logits, we stacked the following activation:
\begin{equation}
\sigma_{k}(\mathrm{z})= 1 + ELU(\mathrm{z}) + \epsilon , \quad\text { with: }  
\end{equation}
\begin{align}
ELU(\mathrm{z})=
\begin{dcases}
\mathrm{z} \quad& \mathrm{if:}\quad \mathrm{z}\geq 0\\
\psi(\mathrm{e}^\mathrm{z}-1) \quad& \mathrm{if:}\quad \mathrm{z}< 0\\
\end{dcases}
\end{align}
where $\psi \in [0.1,0.3]$ and $\epsilon$ is a small number (e.g, 1e-8), to avoid potential NaN during approximate loss computation \cite{10.1007/978-3-319-77583-8_11}.
As regarding the mean outputs, since they do not have particular computational constraints to be addressed, we employ the linear mappings of the conventional MDN form \cite{Bishop:2006:PRM:1162264}.

Then, the architectural form of the neural network must be defined. As a general requirement, to tackle challenging PLF tasks in volatile contexts, the NN must support flexible mappings of arbitrarily complex relationships between the input variables and output distribution parameters.
Various network forms might be considered for such purpose, including feedforward and recurrent architectures \cite{Goodfellow-et-al-2016}. In this work, we exploit a feed-forward form, by providing the past values of the conditioning features as input set over a properly configured time-window. 
Considering two hidden layers of $n_{h_1},n_{h_l} \in \mathbb{Z}^+$ units to lighten notation, the network architecture is mathematically expressed as:
\begin{align}
\begin{aligned}
h_{i}^{(1)} =& f_{i}^{(1)}\left( \sum_{d=1}^{n_x}\omega_{d,i}^{(1)}\mathrm{x}_{n,d} + \omega_{0,i}^{(1)}\right)  \\
h_{j}^{(l)} =& f_{j}^{(l)}\left( \sum_{j=1}^{n_{h_1}}\omega_{i,j}^{(l)}h_{i}^{(1)} + \omega_{0,j}^{(l)}\right)  \\
\mu_{k}=&h_{[\mu_k]}^{(l)} \\
\alpha_{k}=&\frac{\exp \left(h_{[\alpha_k]}^{(l)}\right)}{\sum_{j=1}^{n_k} \exp \left(h_{[\alpha_j]}^{(l)}\right)} \\
\sigma_{k}=&\left(1+ELU\left(h_{[\sigma_k]}^{(l)}\right)+\epsilon \right)
\end{aligned}
\end{align}
where $\omega_{d,i}^{(1)} \in \mathbb{R}^{n_x \times n_{h_1}}$, $\omega_{i,j}^{(l)} \in \mathbb{R}^{n_{h_1} \times n_{h_l}}$, $\omega_{0,i}^{(1)} \in \mathbb{R}^{n_{h_1}}$, $\omega_{0,j}^{(l)} \in \mathbb{R}^{n_{h_l}}$ represent the network weights and biases, $f_{i}^{(1)},f_{j}^{(l)}$ the hidden units activation functions, and $h_{[\mu_k]}^{(l)}, h_{[\sigma_k]}^{(l)}, h_{[\alpha_k]}^{(l)}$ the upper hidden layer partition into the component-wise GMM parameters, respectively. 
Further implementation details are provided in section 3. 
We might remark here that, as the proposed PLF approach is agnostic to the specific conditioning network form employed, the investigation of alternative architectures is envisioned as future extension of the present study.
 
To train MDNs, regularized log-likelihood optimization techniques have been considered within previous studies \cite{8988175}.  Besides having computational problems to be properly tackled (e.g, mode collapse), such approaches do not capture the epistemic uncertainty in the models.  Indeed, point estimates of the model parameters are finally inferred. In the next sections, we address such issues by introducing Bayesian deep learning techniques in our PLF framework.

\subsection{Achieving reliable Bayesian MDNs by approximate inference}
Since standard Bayesian inference methods, commonly exploited for simple regression models and small data contexts, are not feasible for complex neural networks, a lot of research effort has been dedicated in the last years to scalable computation approaches (see e.g., \cite{quality_BDL},\cite{10.1145/3409383} for detailed reviews). 
In particular, relaxed mini-batch versions of the standard Markov Chain Monte Carlo method have been proposed, such as Stochastic Gradient Langevin Dynamics \cite{10.5555/3104482.3104568} and Stochastic gradient Hamiltonian Monte Carlo \cite{pmlr-v32-cheni14}, but still suffering rather slow mixing rate, quite correlated sampling, and lack of convergence guarantee when related strong assumptions are not satisfied \cite{9150658}. Hence, Variational inference (VI) techniques, providing efficient approximations to the intractable posterior via more convenient distributions, are subject of increasing research interest \cite{liberty}. Therefore, we focused on VI to setup the proposed PLF method.   

Specifically, the Bayesian MDN inference task is tackled by minimizing the Kullback-Leibler (KL) divergence (i.e. relative entropy) from the latent posterior, formally expressed as: 
\begin{align}
\begin{aligned}
D_{\rm \scriptscriptstyle  KL}\left(q_\lambda\left(\omega\right) \vert\vert p(\omega|\mathcal{D})\right) &= -\int_\Omega q_\lambda\left(\omega\right)
\log\left(\frac{p\left(\omega\vert \mathcal{D}\right)}{q_\lambda\left(\omega\right)}
\right)d\omega \\
&= -\int_\Omega q_\lambda\left(\omega\right)
\log\left(\frac{p\left(\omega, \mathcal{D}\right)}{q_\lambda \left(\omega\right)}
\right)d\omega + \log p\left(\mathcal{D}\right)
\end{aligned}
\label{KL}
\end{align}
where, $q_\lambda(\omega)$ is the $\lambda$-parameterized variational distribution approximating the posterior distribution $p(\omega|\mathcal{D})$ of the PLF network parameters. \\
As the second term in ~\ref{KL} is constant with reference to the network parameters and since the KL-divergence is positive by definition, it turns out that the first component controls the difference between the target posterior and the variational distribution. This term is often referred to as the Evidence Lower Bound (ELBO) or variational free energy.

Hence, the joint distribution $p\left(\omega, \mathcal{D}\right)$ can be factorized via the Bayes rule, and rearranging the terms, the ELBO is rewritten in the following form:
\begin{align}
\begin{aligned}
{\rm ELBO}\left(\lambda\right) &= \int_\omega q_\lambda\left(\omega\right)
\log\left(\frac{p\left(\omega\right)}{q_\lambda\left(\omega\right)}
\right)d\omega + \int_\Omega q_\lambda\left(\omega\right)
\log p\left(\mathrm{y} \vert \mathrm{x},\omega\right)d\omega\\
&=-D_{\rm \scriptscriptstyle  KL}\left(q_\lambda\left(\omega\right) \vert\vert p(\omega)\right) +
\mathbb{E}_{\omega \sim q_\lambda\left(\omega\right)}\left[ \log p\left(\mathrm{y} \vert \mathrm{x},\omega\right)\right] \\
\lambda^* &=\arg\underset{\lambda}{\min}\left\lbrace -
\mathbb{E}_{\omega \sim q_\lambda\left(\omega\right)}\left[ \log p\left(\mathrm{y} \vert \mathrm{x},\omega\right)\right] + D_{\rm \scriptscriptstyle  KL}\left(q_\lambda\left(\omega\right) \vert\vert p(\omega)\right)\right\rbrace 
\end{aligned}
\label{elbo}
\end{align}

By exploiting such VI framework, the approximated posterior distribution can be estimated through the minimization of the ELBO with reference to the variational parameters $\lambda$. Afterwards, predictive distributions are obtained from trained PLF models by means of the expectation over the posterior integral ~\ref{post_int}, using samples from the variational approximation.

For the target Bayesian MDN model, the predictive distribution can be expressed as follows:
\begin{align}
\begin{aligned}
p(\mathbf{y}|\mathbf{x},\mathcal{D}) &=\int_\Omega p(\mathbf{y}|\mathbf{x},\omega)p(\omega|\mathcal{D})d\omega\approx\mathbb{E}_{\omega \sim q_\lambda\left(\omega\right)}\left[ p(\mathbf{y}|\mathbf{x},\omega)\right] \\
&\approx \frac{1}{\mathcal{M}}\sum_{m=1}^{\mathcal{M}}\sum_{k=1}^{n_k}\alpha_{k}(\mathrm{x},\omega^{(m)}) \phi_{k}(\mathrm{y} | \mathrm{x},\omega^{(m)}),\text { with: } \omega^{(m)} \sim q_\lambda\left(\omega\right)
\end{aligned}
\end{align}
where $\left\lbrace \omega^{(m)}\right\rbrace_{m=1}^\mathcal{M}$ represents a set of samples from the variational posterior.

Notably, the reported VI approach to Bayesian MDN estimation is agnostic to the specific class of distribution. The next step is the specification of the variational class $q_{\lambda}(\omega)$ employed in the PLF model, as detailed in the following subsection.

\subsection{Specification of the variational distribution class}
To achieve a reliable inference process, enabling the adoption of enhanced Bayesian MDN models in practical LF applications, we deployed a Mean Field (MF) variational approximation \cite{10.1145/168304.168306}. Specifically, a factorized Gaussian posterior form is exploited, given by the product of the neural network weights distributions:
\begin{equation}
q_\lambda(\omega)=\prod_{i=1}^{\Omega}\mathcal{N}\left(\omega_{i}; \mu_{\omega_i}, \sigma_{\omega_i}^2 \right) 
\end{equation}
where $\mu_{\omega_i}, \sigma_{\omega_i}^2$ represents the parameters of the $\omega_{i}$-weight approximate distribution in the $\Omega$ space.\\
The rational behind such choice is twofold. On the one hand, MF provides continuous distribution space support - thus enabling approximate sampling around basins -, as opposed to alternative posterior sampling methods \cite{pmlr-v108-farquhar20a}. On the other hand, it has been recently shown that expressive posteriors in function space can be obtained by using simple shallow networks including complex variational families (as e.g., by explicitly modeling correlations between weights via full/structured covariances, etc.) as well as through relatively simpler weight-distributions (as e.g., MF) together with deep network architecture \cite{liberty}. Hence, we followed the latter approach in order to concurrently address posterior representation capabilities, cheaper computational costs and the mapping flexibility (i.e., through hierarchical hidden representations), fundamental to properly infer the articulated relations between the conditioning features and the target electric load distribution. However, as the present study constitute a first step towards the full exploration of Bayesian MDN for PLF, we foresee the investigation and experimental comparison of further approximate inference techniques in future extensions of the present work.

\begin{figure}[h!]
	\centering
	\includegraphics[width=0.4\linewidth]{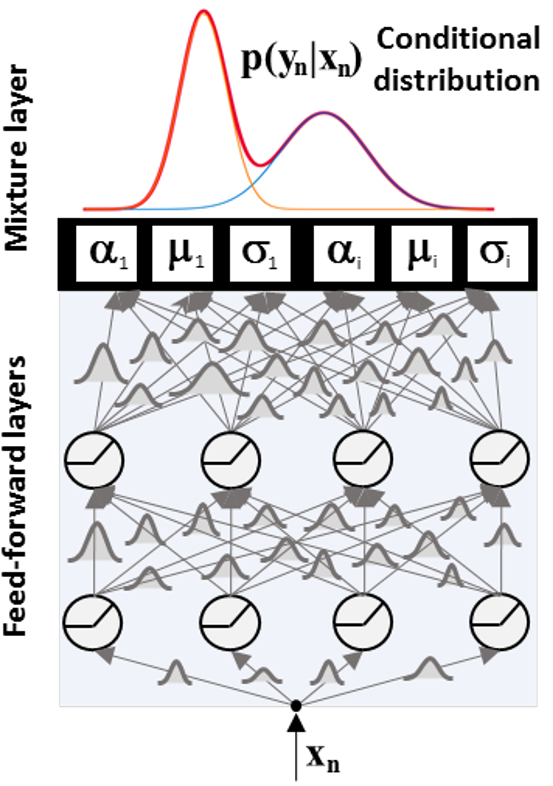}
	\caption{Schematic representation of the Bayesian MDN}
	\label{bmdn}
\end{figure}

A schematic representation of the overall network is reported in Figure~\ref{bmdn}.

Given the MF approximated posterior, different techniques can be employed to obtain estimation of the ELBO gradient with reference to the variational parameters (see e.g., \cite{pmlr-v115-hafner20a} for a detailed review). In this work, we considered the unbiased estimate proposed in \cite{Blundell:2015:WUN:3045118.3045290}, by leveraging on the reparametrization trick:
\begin{equation}
\omega \sim \mathcal{N}(\mu_\lambda, \sigma_\lambda) \Longleftrightarrow \omega=\mu_\lambda + \sigma_\lambda\odot\varepsilon, \text { with: } \varepsilon \sim \mathcal{N}(0,1)
\end{equation}
Hence, sampling can be reframed to neural network weight perturbation using auxiliary Gaussian variables, thus enabling the efficient computation of the posterior parameters with standard back-propagation routines over mini-batches. Besides, flipout provides an efficient mechanism to de-correlate mini-batch gradients through sample-wise pseudo-independent perturbations \cite{wen2018flipout}. The investigation of alternative gradient estimators is left to future developments.

\subsection{Addressing potential mis-specification in Bayesian MDNs via tempering}
In section 2.2, we introduced the ELBO loss, providing a principled approach to minimize the KL-divergence for the variational distribution. As formalized in ~\ref{elbo}, the ELBO is composed by two terms, namely the likelihood expectation and the prior-posterior divergence. The former encourages the learning algorithm to fit parameters values well explaining the available observations. The latter is exploited to induce a kind of Occam’s razor effect, by penalizing complex distributions deviating from the prior settings. Indeed, by employing factorized priors of the form $\mathcal{N}(0,1)$, simpler network parametrizations are enforced to mitigate over-fitting potential.  

However, recent studies have shown that large Bayesian neural networks trained in this fashion can suffer potential over-regularization, which is related to the model mis-specification by the available data \cite{wenzel20a}. 
To avoid this issue, we exploited a safe Bayesian approach, leveraging on a tempered posterior distribution. Formally, the likelihood contribution in ~\ref{post_dist} is scaled as follows:
\begin{equation}
p_\tau(\omega|\mathcal{D})\propto p(\mathcal{D}|\omega)^{1/\tau}p(\omega)
\label{post_dist_temp}
\end{equation}
where $\tau \in \mathbb{R}$ constitute a parameter (a.k.a. temperature) controlling the relative effect of the two components in the overall posterior approximation routine. Indeed, a $\tau$ parameter  lower (greater) than 1 amplifies the likelihood (respectively, the prior) weights in the overall optimization. 

In a Bayesian perspective, tempering incorporates within the inference framework prior beliefs on potential mis-specification of complex neural networks in finite samples conditions \cite{wilson}. Practically, it supports better approximations to the posterior distribution in this settings - by balancing the model capacity to the effective amount of available observations - thus leading to improved predictive performances. 
 
Hence, by introducing the tempered posterior in the inference process, the ELBO minimization problem results in the following weighted form:
\begin{equation}
\lambda^*_\tau =\arg\underset{\lambda_\tau}{\min}\left\lbrace -
\mathbb{E}_{\omega \sim q_\lambda\left(\omega\right)}\left[ \log p\left(\mathrm{y} \vert \mathrm{x},\omega\right)\right] + \tau D_{\rm \scriptscriptstyle  KL}\left(q_\lambda\left(\omega\right) \vert\vert p(\omega)\right)\right\rbrace 
\label{elbo_t}
\end{equation}
Then, we adopted cross-validation to tune the best temperature value in the specific PLF application. 

It is worth nothing that the Bayesian approach deployed within the proposed PLF framework intrinsically provides a facility of practical significance. Indeed, training can be performed end-to-end within the open-source platforms typically employed for conventional maximum likelihood based network training, thus enabling the exploitation of their efficient computational facilities and optimization functions. 
Further details are reported within section 3.

\subsection{Combine posterior basins sampling to improve marginalization}
As introduced in Section 2.1, Bayesian inference of neural PLF models targets the achievement of reliable function space posteriors, thus enabling both accurate forecasting and uncertainty estimations in out of samples conditions. To this end, function space diversity is a critical aspect to be properly addressed \cite{pmlr-v115-hafner20a}. In fact, as deep neural networks are exploited to learn complex mappings given small amounts of observations, quite different settings of the weights can support comparable explanations of the targets (i.e, high likelihood), while still resulting in redundant output functions. Hence, they provide limited contributions to the BMA integral estimation and to the consequent epistemic uncertainty quantification \cite{NEURIPS2019_8558cb40}. 

By investigating the loss landscape of neural networks, it has been recently shown that this issue is strictly related to the effective characterization of multiple modes in the posterior space \cite{fort2020deep}.
Thus, as VI methods target detailed representations concentrated around single basins of attraction (i.e, posterior modes), they could lack in samples heterogeneity, key to proper predictive distribution computation and model generalization \cite{fort2020deep},\cite{zaidi2020neural}.  

Recently, it has been shown that deep neural network ensembles (aka Deep ensembles), traditionally considered as non-Bayesian approaches, perform a kind of approximate marginalization by covering individual samples from different basins (via e.g., single Maximum a Posteriori estimates), reached by randomly initialized trajectories \cite{10.5555/3295222.3295387}. 

Therefore, beyond the VI- approximation reported in previous sections, we included a Deep Ensemble technique within our PLF framework. 
Formally, the function space density is approximated as follows for the MDN model:
\begin{align}
\begin{aligned}
p(\mathbf{y}|\mathbf{x},\mathcal{D}) &\approx \frac{1}{ n_e}\sum_{e=1}^{n_e}\sum_{k=1}^{n_k}\alpha_{k}(\mathrm{x},\omega_e) \phi_{k}(\mathrm{y} | \mathrm{x},\omega_e)
\end{aligned}
\end{align}
where $\left\lbrace \omega_{e}\right\rbrace_{e=1}^{n_e}$ are the parameters of the sub-networks constituting the ensemble. 

\begin{figure}[h!]
	\centering
	\makebox[\linewidth]{
		\includegraphics[width=0.7\linewidth]{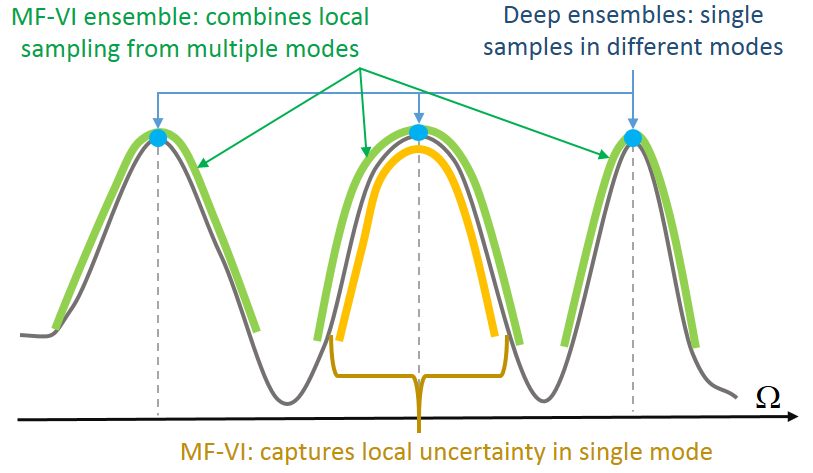}
	}
	\caption{Schematic representation of posterior modes sampling}
	\label{postModes}
\end{figure}

Still, although providing significant contributions to functional heterogeneity, Deep ensembles lack full support in the parameter space and in-mode marginalization of VI techniques \cite{liberty}. In fact, the combination of the strength of both approaches, to marginalize across and within posterior modes, is a promising and open field of research \cite{wilson}.
Such concept is displayed in Figure~\ref{postModes}, taking inspiration from \cite{fort2020deep}.

Therefore, to achieve both local (i.e, single-basins) and global (i.e, multi-mode) approximation capabilities within a unique inference machinery, we introduced an integrated approach based on a Mean Field-Bayesian MDN ensemble. Notably, as ensembles can be executed in parallel during both training and test phases by leveraging on modern GPUs, such functional extension do not introduce critical overheads on the PLF model deployment time.
Specifically, we compute a set of $n_e \in \mathbb{Z}^+$ inference trajectories from random starting conditions, using the Mean Field approximation method reported in previous subsections. Afterwards, we perform a Monte Carlo approximation of the integral by combining samples generated by each ensemble member, as follows:
\begin{equation}
p(\mathbf{y}|\mathbf{x},\mathcal{D}) \approx \frac{1}{\mathcal{M}\cdot n_e}\sum_{m=1}^{\mathcal{M}}\sum_{e=1}^{n_e}\sum_{k=1}^{n_k}\alpha_{k}(\mathrm{x},\omega_e^{(m)}) \phi_{k}(\mathrm{y} | \mathrm{x},\omega_e^{(m)}),\omega_e^{(m)} \sim q_{\lambda_e}\left(\omega_e\right)
\end{equation}

As diversity constitute a standard requirement in conventional network ensembles, several techniques have been proposed in the related literature to foster such characteristic during training (see e.g., \cite{ijcai2020-296} for a review). In this work, considering the highly non-convex nature of the loss at hand, we exploit a straight approach, based on different random initialization and training data shuffles in each component of the ensemble. We envision the investigation of further methods to future extensions.
 
It is worth to remark that, while ensembles are often considered in the machine learning context as a way to enrich the hypothesis space (hence data explanation through models combination), the ensembling mechanism exploited in this work performs a kind of soft model selection \cite{UsingBMA} - i.e., averaging due to the inability to distinguish over weights settings given finite observations -, consistent with the target BMA. The investigation of further ensembling techniques, combining multiple models beyond BMA to enrich the hypothesis space, is left to future extensions.

\subsection{Predictive forecast evaluation criteria}
As observed in \citealp{hong2016probabilistic}, no formally-stated standard exists for PLF, which may impact assessments consistency and comparability between different research studies. In general, as the goal of density forecasting is to infer the latent distribution of the load given the conditioning variables, a proper assessment of the experimental results must consider both the concentration of each prediction around the target and the accuracy of the related uncertainty estimate \citealp{JSSv090i12}. The former requirement, which is often referred to as sharpness, reward models having lower input-dependent variance, i.e., greater observation density. However, it does not address the quality of the uncertainty estimate, fundamental for a proper probabilistic forecasting system, to achieve trustworthiness and consequent adoption in practical applications \cite{pmlr-v80-kuleshov18a}. In fact, the latter requirement, which is often referred to as calibration, focus on the statistical consistency of the predicted distributions \cite{doi:10.1146/annurev-statistics-062713-085831}. Specifically, it considers the coherence between the predicted probabilities and the observed long-run occurrences of events, checked in out-of-samples conditions. 

To achieve reliable PLF systems, such orthogonal objectives must be concurrently optimized, i.e., maximize predictive distribution sharpness subject to calibration \cite{https://doi.org/10.1111/j.1467-9868.2007.00587.x}. Consequently, various summary measurements  - unifying both aspects - have been proposed to correctly rank probabilistic forecasters \citealp{JSSv090i12}. Strictly Proper Scoring Rules (SPSR) are principled tools for such purpose \cite{doi:10.1198/016214506000001437}. In particular, the Continuous Ranked Probability Score (CRPS), a special case of the general energy score, is broadly adopted as a de-facto standard in regression settings, including PLF (see e.g., \citealp{hong2016probabilistic} and references therein). Indeed, CRPS enjoy various appealing features, such as robustness and sensitivity to distances, while rewarding densities around the realizations. We refer to \cite{doi:10.1198/016214506000001437} for a more detailed review and analysis of the mathematical properties. 

Accordingly, we adopted CRPS to evaluate the performances of the probabilistic models. Formally, CRPS is defined as follows:
\begin{equation}
CRPS(P,\mathbf{y})=-\int_{-\infty}^{+\infty}\left[P(z)-\mathbf{1}\left\lbrace z\geq \mathbf{y}\right\rbrace  \right]dz 
\end{equation}
where $P(z)$ denotes the predictive cumulative distribution function (CDF) and $\mathbf{1}\left\lbrace .\right\rbrace$ the indicator function. Under finite first moment of $P(\mathbf{y})$, the CRPS can be expressed in the form:
\begin{equation}
CRPS(P,\mathbf{y})=\mathbb{E}_P|y-\mathbf{y}|-\frac{1}{2}\mathbb{E}_{P,P}|y-y'|
\end{equation}
give independent samples $y,y'$ from the distribution.

Then, by exploiting the empirical approximation to the predictive distribution, CRPS can be operationally computed over each target sample $y_n$ through: 
\begin{equation}
CRPS_N=\frac{1}{N}\sum_{n=1}^{N}\left[ \frac{1}{m}\sum_{i=1}^{m}|y_{n}^i-y_n|-\frac{1}{2m^2}\sum_{i=1}^{m}\sum_{j=1}^{m}|y_{n}^i-y_{n}^j|\right] 
\end{equation}
where $m$, $N$ represents the number of the samples from the predictive distribution and the target dataset size respectively. CRPS is negatively oriented. Hence, the performance of probabilistic forecasters are ranked according to the lowest average score on out-of-sample data.

\section{Applications and results}
In this section we report the experimental verification of the proposed PLF techniques through the application to real case-studies. As observed in \cite{XU2019180}, most of previous works targeted forecasting tasks at aggregation-system level. However, due to the increasing availability of distributed measurements,  thanks to the widespread installation of embedded smart meters, individual LF tasks (e.g., at building/household level) are attracting increasing research interest to capture further dependencies from raw time series and construct hierarchical LF algorithms \cite{hong2016probabilistic}. Despite being still less developed, such fine-grained problems are widely recognized as interesting and complementary PLF benchmarks due to their greater volatility and heterogeneity as compared to the aggregated loads cases \cite{yang2019deep}. Therefore, we considered the latter class of PLF problems to test the proposed approach, adopting the UK-Power Network Smart Meter Energy Consumption dataset \cite{uk-smec} (labelled UK-SMEC) previously employed in \cite{afrasiabi2020deep}. \\
Specifically, the UK-SMEC dataset provide half hourly load measures of 5,567 London house between November 2011 and February 2014, collected during the Low Carbon London project. Interestingly, the aim of the project was to explore novel Dynamic Time of Use (dToU)  energy prices, thus leading to particularly volatile load patterns during the night hours, as opposed to conventional household consumption under fixed hourly price conditions.  
Following \cite{afrasiabi2020deep}, we randomly selected a subset of households, reported in Table~\ref{UK_id}. 

\begin{table}[ht]
	\small
	\centering
	\caption{Identification code of the households in UK-SMEC dataset.}
	\makebox[\linewidth]{
		\begin{tabular}[t]{lcccccccc}
			\toprule
			&H\#1 &H\#2 &H\#3 &H\#4\\
			\hline
			ID   &MAC005041 &MAC004970 &MAC004902 &MAC004897\\
			\midrule
			&H\#5 &H\#6 &H\#7 &H\#8\\
			\hline
			ID   &MAC004866 &MAC001477 &MAC000415 &MAC000032\\
			\bottomrule
		\end{tabular}
		\label{UK_id}
	}
\end{table}%

\begin{figure}[b!]
	\centering
	\makebox[\linewidth]{
	    \includegraphics[width=1.1\linewidth]{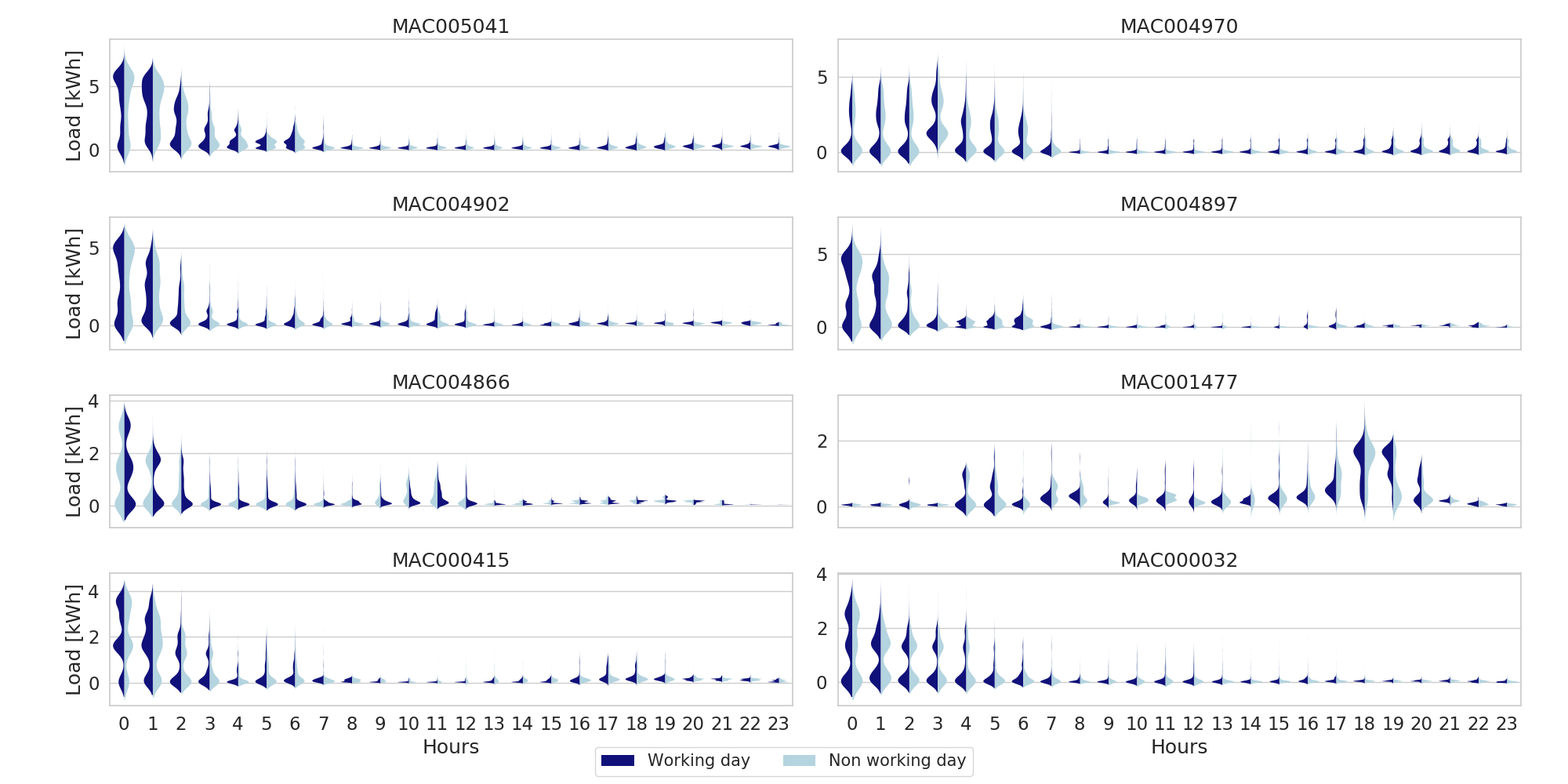}
    }
	\caption{Hourly load distributions of UK-SMEC households}
	\label{UK_hour}
\end{figure}

\begin{figure}[h!]
	\centering
	\makebox[\linewidth]{
	    \includegraphics[width=1.1\linewidth]{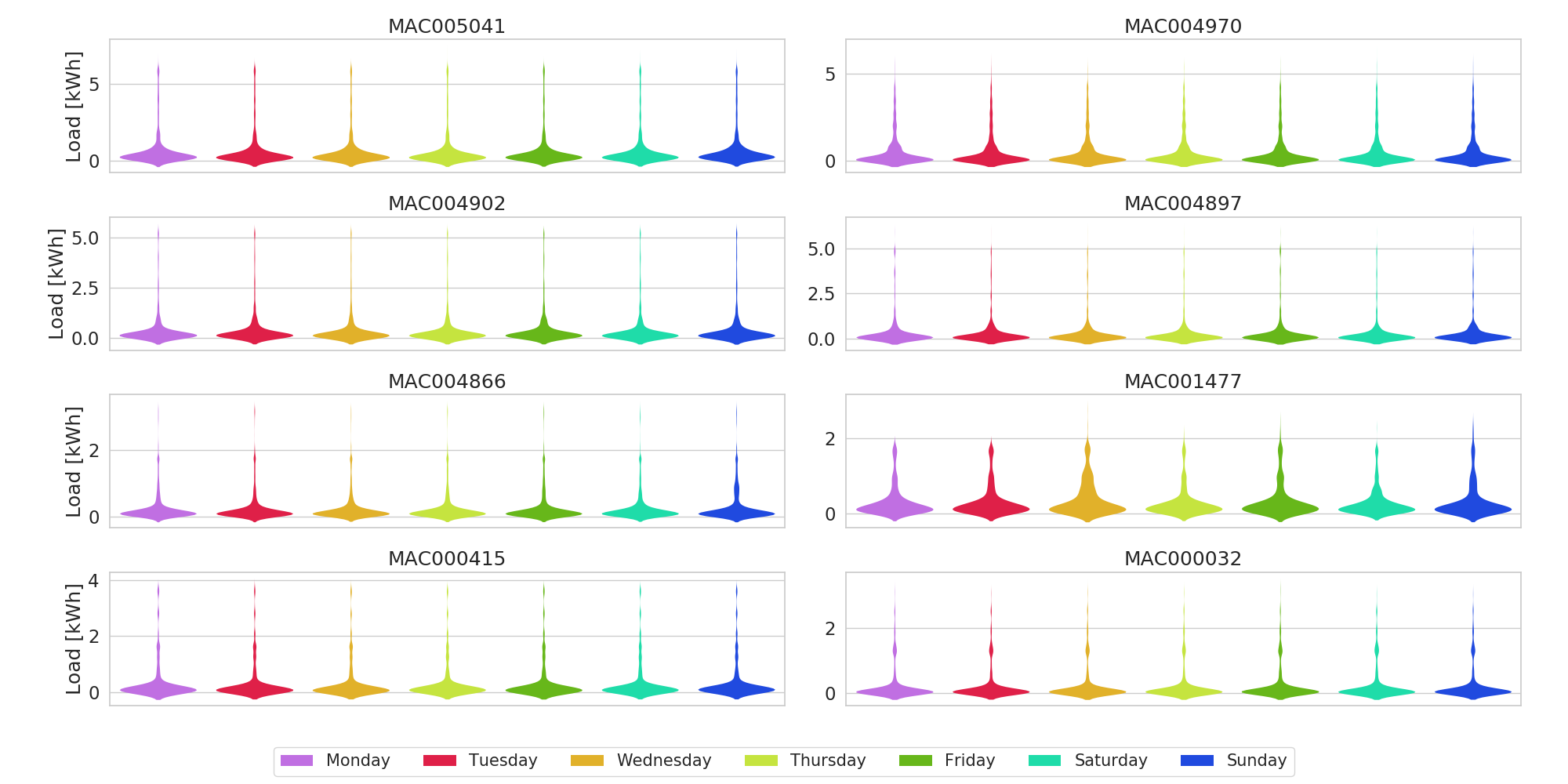}
	}
	\caption{Daily load distributions of UK-SMEC households}
	\label{UK_day}
\end{figure}

\begin{figure}[h!]
	\centering
	\makebox[\linewidth]{
	    \includegraphics[width=1.1\linewidth]{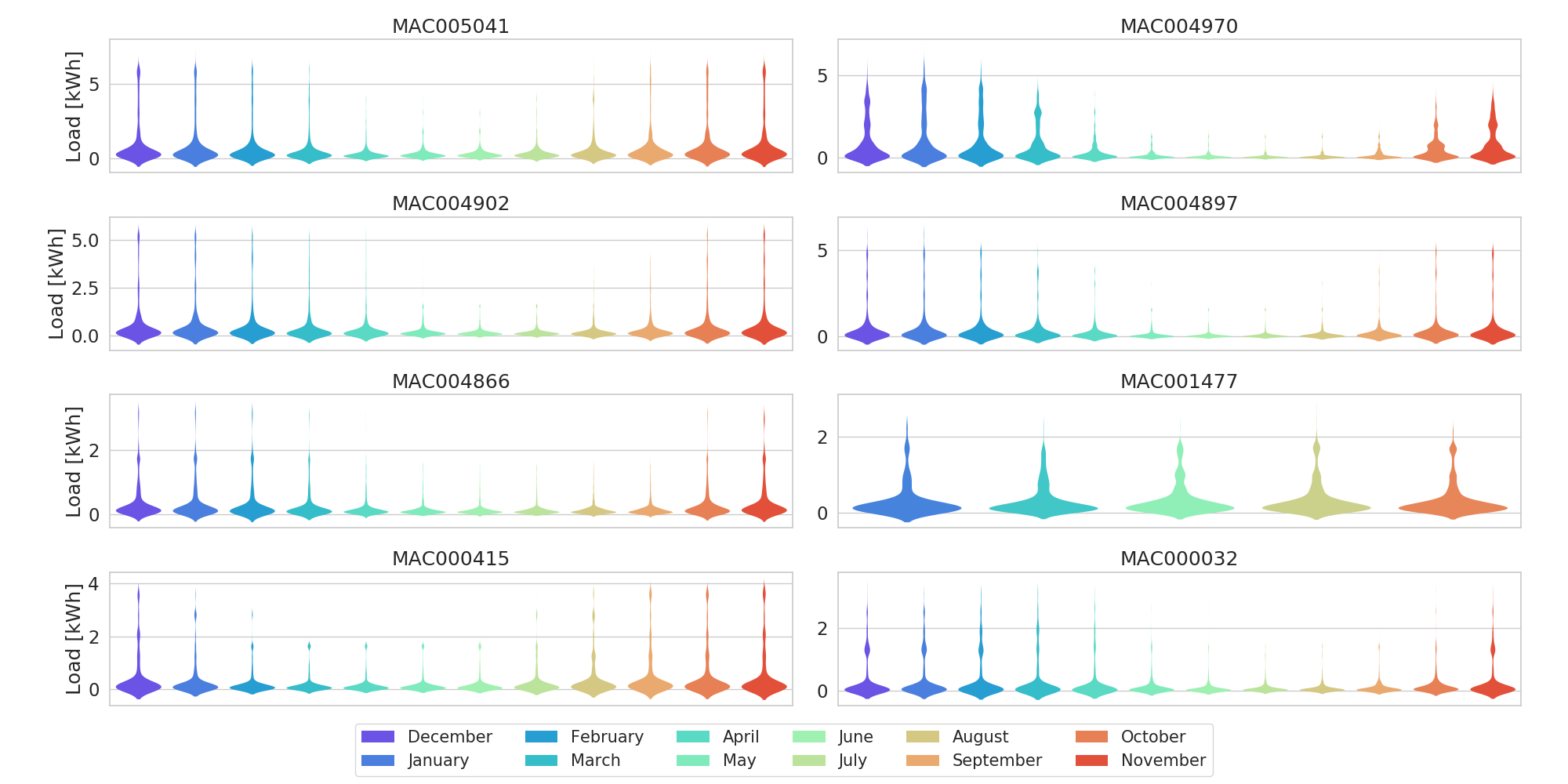}
	}
	\caption{Monthly load distributions of UK-SMEC households}
	\label{UK_mon}
\end{figure}

\begin{figure}[h!]
	\centering
	\includegraphics[width=1.0\linewidth]{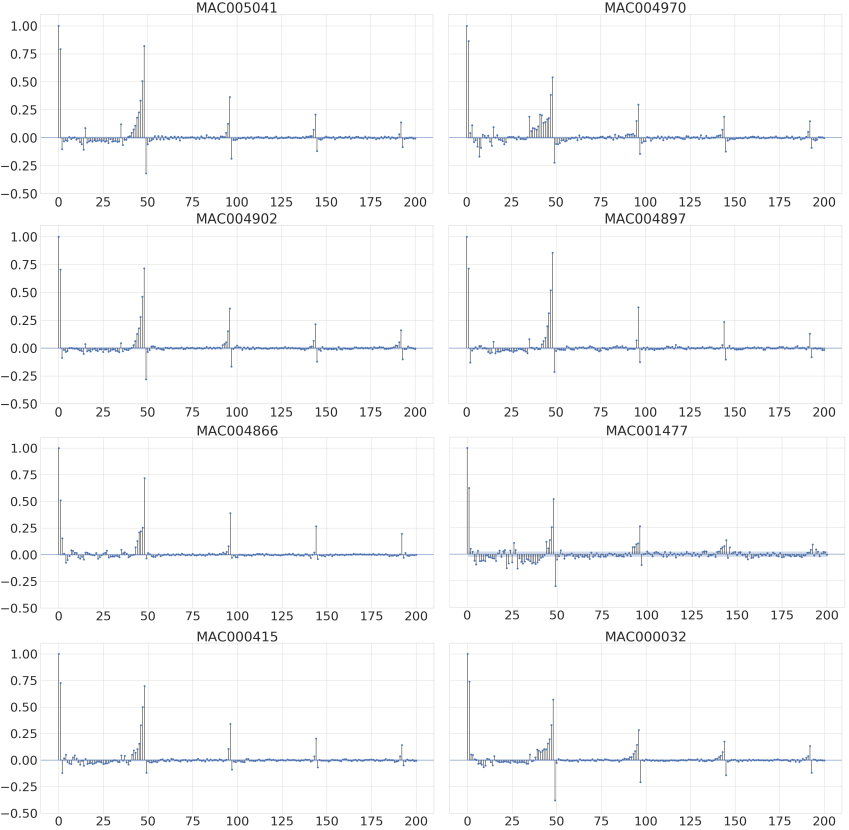}
	\caption{PACF of loads sequences (on half-hour lags) in UK-SMEC households}
	\label{UK_PACF}
\end{figure}

The major characteristics of the dataset are visualized in Figure~\ref{UK_hour}-~\ref{UK_PACF}, reporting the daily and hourly distributions, and Partial Auto-Correlation Functions (PACF). Table~\ref{UK_summ} summarizes principal descriptive statistics of the marginal distribution in each unit.\\
\begin{table}[ht]
	\small
	\centering
	\caption{Summary statistics of UK-SMEC dataset [kWh].}
	\makebox[\linewidth]{
		\begin{tabular}[t]{lcccccccc}
			\toprule
			&H\#1 &H\#2 &H\#3 &H\#4 &H\#5 &H\#6 &H\#7 &H\#8\\
			\midrule
			mean   &0.704 &0.561 &0.453 &0.370 &0.276 &0.341 &0.357 &0.267\\
			std    &1.182 &0.959 &0.881 &0.846 &0.467 &0.423 &0.666 &0.526\\
			25\%   &0.166 &0.040 &0.090 &0.031 &0.055 &0.082 &0.044 &0.033\\
			50\%   &0.269 &0.103 &0.167 &0.089 &0.095 &0.177 &0.120 &0.059\\
			75\%   &0.578 &0.617 &0.293 &0.229 &0.201 &0.357 &0.219 &0.151\\
			\bottomrule
		\end{tabular}
		\label{UK_summ}
	}
\end{table}%
It is worth noting that, for most cases, the consumption pattern is visibly influenced by the dToU price, characterized by load consumption concentrated in lower price periods (i.e., nigh-time). A single exception is MAC001477, characterized by a typical fixed price form. Moreover, this household include a quite minor number of samples (approximately 5 months), thus constituting a interesting testing scenario regarding further lower samples regimes. 
Besides, each sub-case has specific characteristics, both in terms of scale, patterns and dispersion.

The scope of the benchmark is to perform day-ahead predictions, i.e., forecast the load for each hour of the next day given the conditioning variables available till the current day. As input features from the available variables in the UK-SMEC dataset, considering \cite{afrasiabi2020deep} and the major peaks visible in the PACF, we adopted the two lags t-24h, t-48h besides month/weekday/hour indicators.
We remark that further improvements might be obtained by including more specific conditioning variable within the models. However, as the aim of this work is to compare PLF techniques under consistent conditions, we leave such investigation to future extensions, e.g., by exploiting automatic feature selection mechanisms within the forecasting framework.    
Following the characterization of the features set, we structured the overall data-sets into a supervised learning form by applying a sliding window, thus extracting evenly spaced batches ordered according to the original time series. 
Afterwards, we split the samples into training, validation and test subsets by a 70\%/15\%/15\% decomposition. 

To achieve well conditioned problems during training, we performed samples standardization by subtracting the mean and scaling to unitary variance both inputs and targets. As common, model outputs are re-conducted to the original order (i.e., by inverting the scaling procedure) for subsequent forecasting performance assessment.  
 
Afterwards, we proceed with the specification of the neural networks configurations. In general terms, a huge set of hyper-parameters might be experimentally analyzed, including architectural layers shape, stochastic training algorithm set, mini-batch size,  training epoch, etc. Considering the scope of the present work – i.e., the investigation of enhanced  NN based PLF independently from (i.e. given) specific model configuration -,  as well as the computational budget required, we constrain the search space to a reduced dimension by fixing several potential hyper-parameters to conventional settings and adopt a straight grid search in cross validation. Nevertheless, we envision a more extensive exploration over the hyperparameters space, e.g., through the integration of advanced search algorithms  (as e.g., bayesian optimization based) to future extensions of the present work. 

Specifically, we adopted feed-forward hidden layers with Rectified Linear Unit (ReLU) activations, trained by means of the ADAM  algorithm with a learning rate of 1e-3, particularly tailored for noisy and sparse gradients \cite{adam}. The maximum amount of learning epochs have been configured to 10000, including a patience callback of 50 epochs to interrupt the procedure once the validation performance stop decreasing, thus reducing training time. Mini-batch size has been set to 512 samples, constituting a reasonable settlement to achieve suitable gradient estimation and computational load. Random training data shuffling has been performed before each run. Test set configurations are chosen by comparing validation performances reached by five random executions.  Xavier-uniform initializations have been employed for deterministic layers, while zero-mean unit-variance priors have been considered for Bayesian parameters. By cross-validation, we did not observe sensible variation of performances for network architecture above three hidden layers of 100 neurons each, still representing a consistent configuration to support epistemic uncertainty estimation by the variational approximation, as explained in section 2. As regarding the parameterized Gaussian Mixture in the output layer, we determined three components as a reasonable choice for test purpose. Posterior temperature has been tuned to 1e-2.
As suggested in \cite{10.5555/3295222.3295387}, we adopted a relatively small ensemble size, combining 5 networks trained in parallel but cross-validated concurrently to investigate overall convergence. The investigation of alternative learning approaches (e.g., by considering cross-validation performances component-wise  over a larger setting and selecting the best convergences to improve ensemble performance) is outside the scope of the present work and left to future extensions. For consistency, the same configuration has been maintained for both deterministic and variational network parametrizations.   
For a fair comparison, we apply a standard L2-norm regularizer (with penalty 1e-2) to the deterministic network layers, besides early stopping, to mitigate overfitting.
 
To deploy the neural networks, we employ the Tensorflow-2.3 open source framework \cite{TF} and the Tensorflow Probability package \cite{tfp}, providing various utilities for probabilistic modeling including statistical distributions, sampling functions, specialized layers, Kullback–Leibler divergence computation, etc.

As first baseline, we adopt the state of the art PLF method of \cite{li2019combining} reported in section 1, combining best performing techniques including Quantile Regression Neural Networks, Gaussian Process Regression and Quantile Regression Gradient Boosting (hereafter labelled QRNN-QRGB-GP). Besides, we consider also a deterministic Mixture Density Network form, as exploited in \cite{afrasiabi2020deep}, to investigate the specific performance gains given by the introduction of Bayesian inference approaches. 
 
To achieve detailed quantitative assessments of the benefits provided by each enhanced components in the proposed PLF approach, we first analyze forecasting performances obtained through:
\begin{enumerate}
	\item a regularized deterministic NN trained in a conventional maximum a posteriori fashion (labelled GaussNN-homo), leading to a conditional mean prediction followed by a validation set estimation of the overall standard deviation, as detailed in section 2.  
	\item a deterministic MDN including a single component (labelled GaussNN-hete), thus leading to an heterosckedastic Gaussian extension of the model in bullet 1. 
	\item a deterministic MDN including multiple components to infer conditional distribution of general form (labelled DetMDN)
\end{enumerate} 
To avoid biased results, we adopt coherent networks and training configurations in each setup.
Hence, we investigate the specific benefits given by a more detailed input feature-conditioned characterization of the aleatoric uncertinity counterpart (i.e., from simpler Gaussian to Mixture Density forms). Afterwards, we experiment the different bayesian MDN approximations, namely single MDN-variational inference (labelled BayMDN-VI), deep ensemble (labelled BayMDN-DE) and ensemble of MDN-variational inference (labelled BayMDN-DEVI), thus leading to both intra and multi-mode approximate posterior sampling.    
       
CRPS indicators are computed over 500 independent random samples from the probabilistic models. Sampling if performed first on the lower feed-forward network layers, and then at the stacked GMM output. Clearly, deterministic layers provides equivalent parametrizations to the mixture density distribution given the input features. Samples from GaussNN-homo models are obtained using specific Gaussian distributions with mean given by the network output and validation set standard deviation. In the ensemble, sampling is performed by the overall mixture aggregation by uniformly weighting the components, as detailed in section 2.

The obtained test set results are summarized in Table~\ref{UK_crps}. 

\begin{table}[ht]
	\small
	\centering
	\caption{Overall CRPS performance on UK-SMEC Test set.}
	\makebox[\linewidth]{
	\begin{tabular}[t]{lcccccccc}
		\toprule
		&H\#1 &H\#2 &H\#3 &H\#4 &H\#5 &H\#6 &H\#7 &H\#8\\
		\toprule
		QRNN-QRGB-GP &0.2922 &0.3250 &0.4111 &0.2237 &0.3007 &0.2171 &0.2330 &0.2205\\
		\midrule
		GaussNN-homo &0.1485 &0.2261 &0.1682 &0.0969 &0.1285 &0.0935 &0.1136 &0.1100\\
		GaussNN-hete &0.1412 &0.2054 &0.1526 &0.0846 &0.1122 &0.0783 &0.0989 &0.1013\\
		DetMDN       &0.1356 &0.1965 &0.1437 &0.0753 &0.1037 &0.0734 &0.0951 &0.0896\\
		\midrule
		BayMDN-VI    &0.1360 &0.1949 &0.1412 &0.0746 &0.1019 &0.0696 &0.0918 &0.0902\\
		BayMDN-DE    &0.1331 &0.1955 &0.1430 &0.0747 &0.1020 &0.0722 &0.0932 &0.0890\\
		BayMDN-DEVI  &\textbf{0.1328} &\textbf{0.1943} &\textbf{0.1405} &\textbf{0.0726} &\textbf{0.0999} &\textbf{0.0685} &\textbf{0.0905} &\textbf{0.0864}\\
		\bottomrule
	\end{tabular}
    }
    \label{UK_crps}
\end{table}

We observe that a more detailed characterization of the aleatoric uncertainty (from simpler Gaussian to general conditioning distribution) already provides sensible performance improvements. This is more evident in Table~\ref{UK_crps_perc}, reporting the incremental performance improvements starting from the homoskedastic Gaussian network configuration.  

\begin{table}[ht]
	\small
	\centering
	\caption{Performance improvements [\%] wrt GaussNN-homo}
	\makebox[\linewidth]{
		\begin{tabular}[t]{lcccccccc}
			\toprule
			&H\#1 &H\#2 &H\#3 &H\#4 &H\#5 &H\#6 &H\#7 &H\#8\\
			\toprule
			GaussNN-hete &4.92  &9.16  &9.27  &12.69 &12.68 &16.26 &12.94 &7.91\\
			DetMDN       &8.69  &13.09 &14.57 &22.29 &19.30 &21.50 &16.29 &18.55\\
			\midrule
			BayMDN-VI    &8.42  &13.80 &16.05 &23.01 &20.70 &25.56 &19.19 &18.00\\
			BayMDN-DE    &10.37 &13.53 &14.98 &22.91 &20.62 &22.78 &17.69 &19.09\\
			BayMDN-DEVI  &\textbf{10.57} &\textbf{14.06} &\textbf{16.47} &\textbf{25.08} &\textbf{22.26} &\textbf{26.74} &\textbf{20.33} &\textbf{21.45}\\
			\bottomrule
		\end{tabular}
	}
	\label{UK_crps_perc}
\end{table}

The substantial gap between MDNs and QRNN-QRGB-GP, as observed also in \cite{afrasiabi2020deep}, is mainly due to the higher volatility at single household scale compared to the regional level considered in \cite{li2019combining}.
Actually, the specific extent depends on the characteristics of the dataset at hand, i.e, requiring PLF models with enhanced representation capabilities to properly capture the intrinsic stochasticity. 

The Bayesian MDN models achieve best performances across all the datasets with reference to the conventional MDN and the QRNN-QRGB-GP method. Since the developed conventional-MDN and Bayesian-MDN architectures shares the same settings regarding aleatoric uncertainty estimation, the observed performance gain is related to the introduction of the Bayesian framework into the MDN model, supporting parameters uncertainty integration beyond regularization. 
In general, BayMDN-DEVI works better than single BayMDN-VI, thus showing the advantage of including different posterior modes to compute the predictive distribution. We observe a unique case where a single BayMDN-VI worked slightly better than the BayMDN-DEVI. Such effect could be related to a particular optimal solution reached during learning with reference to the ensemble components on average. Indeed, diverse runs of the training algorithms usually result in small random fluctuations in final performances, depending on the starting conditions and consequent minimizers reached by the solver within the complex loss landscape. Besides, we have found MDNs to be sensible to poor random initialization, particularly due to their complex parametrization. Notably, such issue is mitigated by the averaging effect induced by the proposed Bayesian training techniques. On the other hand, we notice that proper execution of variational inference is strongly impacted by learning algorithm and hyper-parameters tuning (e.g., network configuration, temperature, stopping conditions, etc.), thus requiring particular attention during cross-validation. Conversely, deep ensembles result more robust, which is indeed expected due to its capacity to average out eventual poor local solutions. 
We observe such facility to be particularly relevant on the some sub-case in the dataset (e.g., H\#1), which could be explained by the intrinsic balance between aleatoric and epistemic uncertainty requirements, as well as the related impact on the network loss landscape and stochastic convergences. While outside the scope of the present study, such issues constitute  interesting directions of future research, e.g., by introducing further mechanisms to foster properly heterogeneous posterior modes, improved initialization, combination of further Bayesian inference machineries, enhanced automatic hyper-parameter tuning, etc. 

Nevertheless, we do not consider probabilistic forecasting performance improvements as the major outcome of Bayesian deep learning for LF. In our view, the major strength of the proposed approach resides in the provided predictive distribution of future electric loads, extending conventional point, interval and quantile regressions while including also the contribution of the uncertainties of the model parameters. Such overall probabilistic description can be exploited in multiple ways, depending on the specific application requirements. For instance, stochastic and multi-scenario analysis can be performed (thus enabling improved supply side scheduling, generators commitment optimization, detailed risk assessments, etc.) thanks to the availability of samples from the conditional distribution. 

\begin{figure}[t!]
	\centering
	\includegraphics[width=0.8\linewidth]{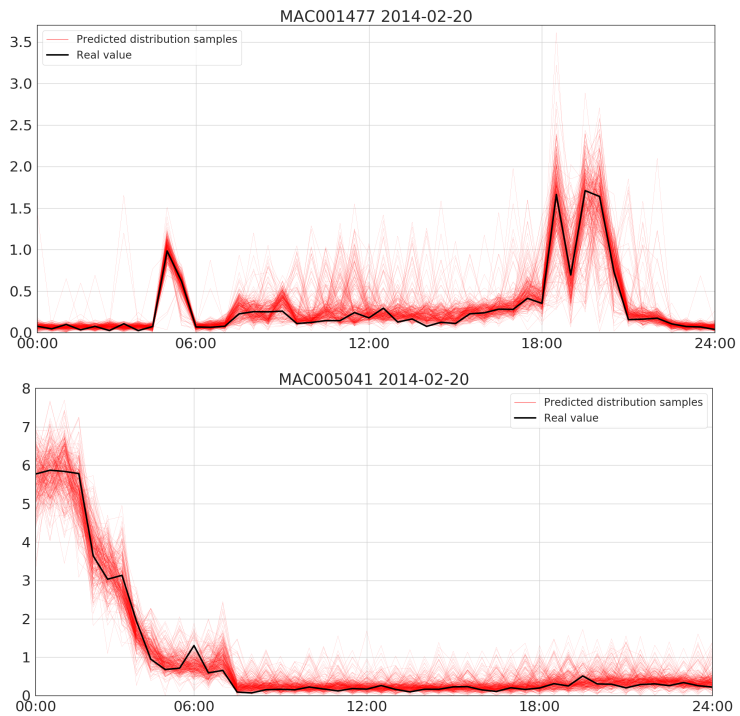}
	\caption{Predicted distribution samples over instances from  UK-SMEC testset}
	\label{pred_samp}
\end{figure}
\begin{figure}[t!]
	\centering
	\includegraphics[width=0.8\linewidth]{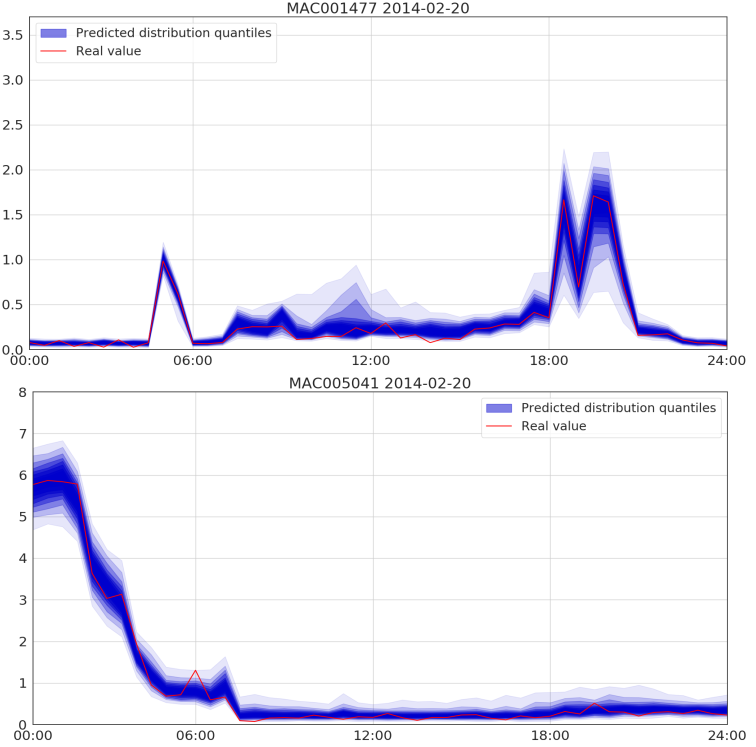}
	\caption{Predicted distribution quantiles over instances from UK-SMEC testset}
	\label{dist_quantiles}
\end{figure}

\begin{figure}[h]
	\centering
	\includegraphics[width=0.8\linewidth]{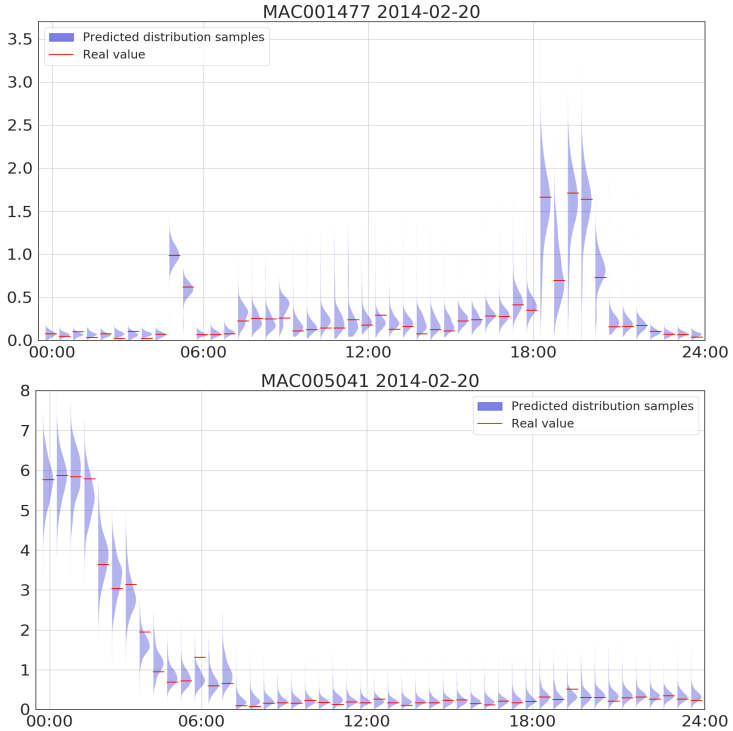}
	\caption{Predicted distributions over instances from UK-SMEC testset}
	\label{predDist_MACs}
\end{figure}

Such facility is displayed in Figure~\ref{pred_samp}, showing a set of random outputs from the probabilistic models over the predicted horizon. Furthermore, hour-specific information, including statistics, intervals, etc., can be straightly extracted to provide further user interpretable summaries. Figure~\ref{dist_quantiles} reports examples of predicted distribution quantiles over different test set conditions, with reference to the actual load, while Figure~\ref{predDist_MACs} includes instances of out-of-samples probability distributions. Visibly, hour-specific uncertainty patterns (i.e, less/more sharped) are obtained, which depends on the feature specific volatility level (e.g., lower/higher peak consumption times) and the distance from the observations accessible during inference. Besides, the actual loads resulted properly covered by the predicted distribution, including times with higher volatility. Visibly, UK-SMEC includes both long tails and skewed patterns in the hourly distributions, as common for fine-grained load series. To support further detailed representation in such sharp settings, we envision the integration of more concentrated densities in the mixture layer within future developments, e.g., considering Laplace components.  

To provide deeper insights, Tables~\ref{crps_hour_uk1}-~\ref{crps_day_uk8} (reported in appendix) include a further detailed decomposition of the networks CRPSs obtained over the test sets, considering hour/day specific calculations. 
We observed slight variations in the CRPSs at specific operating conditions level between models providing consistent prediction performances, which could be related to different parametrizations occurred under limited or sparse observations. In-depth investigations of the latent dynamics behind such observations are left to future developments, e.g., by training and comparing specific network configurations over different operating conditions, include data augmentation techniques or considering further (e.g, CRPS-based) combination approaches between submodels. 
 
\begin{table}[h]
	\small
	\centering
	\caption{Hourly CRPS performance on UK-SMEC-H\#1 Test set.}
	\makebox[\linewidth]{
		\begin{filecontents*}{figures/tables/hourlyCRPS1.csv}
H,GNN-homo,GNN-hete,DetMDN,BMDN-VI,BMDN-DE,BMDN-DEVI
0,0.1836,0.1845,0.174,0.2245,0.1731,0.1983
1,0.3532,0.3158,0.3152,0.3191,0.3166,0.3124
2,0.2750,0.2691,0.2712,0.2673,0.2665,0.2631
3,0.2221,0.2259,0.2262,0.2205,0.216,0.2139
4,0.1635,0.1632,0.1667,0.1593,0.1528,0.1505
5,0.1676,0.1652,0.1677,0.1564,0.1549,0.1489
6,0.2807,0.2678,0.2677,0.2538,0.2605,0.2533
7,0.1014,0.089,0.063,0.0688,0.0618,0.0649
8,0.0879,0.0741,0.0544,0.0579,0.0547,0.0564
9,0.1080,0.0966,0.0891,0.0895,0.0897,0.0893
10,0.0998,0.0901,0.0857,0.0848,0.0849,0.084
11,0.1075,0.0963,0.0923,0.0917,0.0924,0.091
12,0.1035,0.0965,0.0936,0.0936,0.0941,0.0927
13,0.1155,0.1085,0.1037,0.1036,0.1038,0.103
14,0.1097,0.1067,0.1032,0.1033,0.1035,0.1032
15,0.1210,0.1208,0.1193,0.1196,0.1201,0.1196
16,0.1162,0.1148,0.1122,0.1115,0.1122,0.1108
17,0.1390,0.1389,0.1386,0.1368,0.138,0.136
18,0.1301,0.1277,0.1253,0.1225,0.1251,0.1221
19,0.1113,0.1089,0.1003,0.1003,0.1006,0.1011
20,0.1065,0.1035,0.0937,0.0943,0.0936,0.0939
21,0.1233,0.1151,0.1072,0.1057,0.1068,0.1065
22,0.1260,0.1131,0.1022,0.0995,0.0974,0.0978
23,0.1131,0.0974,0.0829,0.0823,0.0778,0.0766
\end{filecontents*}
\pgfplotstabletypeset[
		col sep=comma,
		string type,
		every head row/.style={%
			before row={\hline
			},
			after row=\hline
		},
		every last row/.style={after row=\hline},
		columns/name/.style={column name=Name, column type=l},
		columns/surname/.style={column name=Surname, column type=l},
		columns/age/.style={column name=Age, column type=c},
		]{figures/tables/hourlyCRPS1.csv}
	}
	\label{crps_hour_uk1}
\end{table}

\begin{table}[h]
	\small
	\centering
	\caption{Hourly CRPS performance on UK-SMEC-H\#2 Test set.}
	\makebox[\linewidth]{
		\begin{filecontents*}{figures/tables/hourlyCRPS2.csv}
H,GNN-homo,GNN-hete,DetMDN,BMDN-VI,BMDN-DE,BMDN-DEVI
0,0.2736,0.2467,0.2394,0.2404,0.239,0.2396
1,0.2312,0.2133,0.2092,0.2095,0.2071,0.2081
2,0.2229,0.2131,0.2021,0.2017,0.2009,0.203
3,0.3534,0.3026,0.3137,0.3116,0.3083,0.3119
4,0.5341,0.456,0.4467,0.4454,0.4456,0.4436
5,0.4431,0.3799,0.368,0.3588,0.3679,0.3617
6,0.5025,0.4166,0.4052,0.3911,0.4007,0.3911
7,0.1618,0.1558,0.1473,0.1464,0.1476,0.1451
8,0.1284,0.1292,0.11,0.1073,0.1105,0.1066
9,0.1553,0.1495,0.1418,0.1423,0.1412,0.14
10,0.1378,0.1343,0.1257,0.1239,0.1245,0.1234
11,0.1371,0.1332,0.127,0.1239,0.1256,0.1239
12,0.1263,0.1243,0.1147,0.1127,0.1135,0.1132
13,0.1436,0.1391,0.1253,0.1226,0.1245,0.124
14,0.1486,0.1455,0.129,0.1281,0.1289,0.1273
15,0.1509,0.1476,0.1299,0.1301,0.1299,0.1296
16,0.1707,0.1632,0.1462,0.1467,0.146,0.1461
17,0.1739,0.1654,0.1538,0.1561,0.1523,0.1519
18,0.1773,0.1681,0.162,0.1619,0.1626,0.1615
19,0.1971,0.1816,0.1752,0.1738,0.1744,0.1733
20,0.2221,0.1991,0.1914,0.1956,0.1915,0.1945
21,0.2022,0.1831,0.1789,0.1793,0.1774,0.178
22,0.2298,0.2032,0.1989,0.1957,0.1978,0.1953
23,0.2079,0.1835,0.1794,0.176,0.1775,0.1749
\end{filecontents*}
\pgfplotstabletypeset[
		col sep=comma,
		string type,
		every head row/.style={%
			before row={\hline
			},
			after row=\hline
		},
		every last row/.style={after row=\hline},
		columns/name/.style={column name=Name, column type=l},
		columns/surname/.style={column name=Surname, column type=l},
		columns/age/.style={column name=Age, column type=c},
		]{figures/tables/hourlyCRPS2.csv}
	}
	\label{crps_hour_uk2}
\end{table}

\begin{table}[h]
	\small
	\centering
	\caption{Hourly CRPS performance on UK-SMEC-H\#3 Test set.}
	\makebox[\linewidth]{
		\begin{filecontents*}{figures/tables/hourlyCRPS3.csv}
H,GNN-homo,GNN-hete,DetMDN,BMDN-VI,BMDN-DE,BMDN-DEVI
0,0.2763,0.2507,0.2395,0.2512,0.2345,0.2496
1,0.3791,0.3517,0.3692,0.3587,0.3591,0.3537
2,0.4750,0.4165,0.4129,0.4155,0.4187,0.4109
3,0.3370,0.3276,0.3218,0.2963,0.318,0.2952
4,0.2714,0.2637,0.2507,0.2353,0.2466,0.236
5,0.3024,0.2829,0.2835,0.2863,0.2804,0.2835
6,0.2966,0.2759,0.2706,0.2734,0.2717,0.2682
7,0.1208,0.1193,0.0588,0.0553,0.0615,0.0536
8,0.1140,0.114,0.1086,0.1056,0.1089,0.1058
9,0.1054,0.1079,0.0888,0.0885,0.0905,0.0878
10,0.1609,0.1595,0.1416,0.1407,0.1409,0.1405
11,0.1533,0.1541,0.1543,0.1526,0.1513,0.157
12,0.2197,0.2176,0.2222,0.1955,0.2233,0.2001
13,0.0789,0.0668,0.053,0.0532,0.0529,0.0528
14,0.0732,0.0506,0.0443,0.0446,0.0444,0.044
15,0.0673,0.0465,0.0443,0.0446,0.0446,0.0445
16,0.0700,0.0509,0.0429,0.0441,0.0435,0.0449
17,0.0862,0.0769,0.0763,0.0761,0.0763,0.0759
18,0.0760,0.0647,0.0563,0.0593,0.0566,0.0583
19,0.0642,0.0524,0.0377,0.0417,0.0376,0.0405
20,0.1062,0.0958,0.0825,0.0796,0.0823,0.0799
21,0.0662,0.052,0.0395,0.0432,0.0402,0.0429
22,0.0758,0.0384,0.0352,0.0338,0.0348,0.0334
23,0.0709,0.0345,0.0226,0.0222,0.0224,0.022
\end{filecontents*}
\pgfplotstabletypeset[
		col sep=comma,
		string type,
		every head row/.style={%
			before row={\hline
			},
			after row=\hline
		},
		every last row/.style={after row=\hline},
		columns/name/.style={column name=Name, column type=l},
		columns/surname/.style={column name=Surname, column type=l},
		columns/age/.style={column name=Age, column type=c},
		]{figures/tables/hourlyCRPS3.csv}
	}
	\label{crps_hour_uk3}
\end{table}

\begin{table}[h]
	\small
	\centering
	\caption{Hourly CRPS performance on UK-SMEC-H\#4 Test set.}
	\makebox[\linewidth]{
		\begin{filecontents*}{figures/tables/hourlyCRPS4.csv}
H,GNN-homo,GNN-hete,DetMDN,BMDN-VI,BMDN-DE,BMDN-DEVI
0,0.2028,0.1994,0.1785,0.18,0.1751,0.1835
1,0.1561,0.1866,0.16,0.1549,0.155,0.1593
2,0.1675,0.1783,0.1766,0.1723,0.1768,0.1681
3,0.0975,0.0965,0.0865,0.0885,0.0871,0.0848
4,0.0746,0.0832,0.071,0.0669,0.0723,0.0642
5,0.1380,0.1361,0.1332,0.1325,0.1311,0.1298
6,0.2358,0.2217,0.2158,0.2136,0.2137,0.2107
7,0.0575,0.0294,0.0252,0.019,0.0219,0.014
8,0.0603,0.0472,0.0445,0.0455,0.045,0.0452
9,0.0542,0.0296,0.0176,0.0197,0.0186,0.0188
10,0.0799,0.0618,0.0511,0.0518,0.0515,0.0517
11,0.0613,0.0429,0.036,0.0357,0.0354,0.0363
12,0.0894,0.0696,0.0709,0.0678,0.0715,0.0647
13,0.0917,0.0712,0.061,0.0596,0.0612,0.057
14,0.0618,0.0459,0.0281,0.0309,0.0274,0.028
15,0.0613,0.0531,0.0277,0.0319,0.0279,0.0267
16,0.1091,0.1053,0.0947,0.0985,0.0949,0.0962
17,0.1972,0.1795,0.1813,0.1657,0.1812,0.1539
18,0.0606,0.049,0.0398,0.0389,0.0397,0.0388
19,0.0534,0.0303,0.0246,0.0262,0.0245,0.0248
20,0.0527,0.0255,0.0229,0.0244,0.0222,0.0234
21,0.0523,0.0231,0.0173,0.0214,0.0175,0.02
22,0.0564,0.0332,0.0313,0.0304,0.0308,0.0301
23,0.0566,0.0346,0.0147,0.0166,0.0141,0.0157
\end{filecontents*}
\pgfplotstabletypeset[
		col sep=comma,
		string type,
		every head row/.style={%
			before row={\hline
			},
			after row=\hline
		},
		every last row/.style={after row=\hline},
		columns/name/.style={column name=Name, column type=l},
		columns/surname/.style={column name=Surname, column type=l},
		columns/age/.style={column name=Age, column type=c},
		]{figures/tables/hourlyCRPS4.csv}
	}
	\label{crps_hour_uk4}
\end{table}

\begin{table}[h]
	\small
	\centering
	\caption{Hourly CRPS performance on UK-SMEC-H\#5 Test set.}
	\makebox[\linewidth]{
		\begin{filecontents*}{figures/tables/hourlyCRPS5.csv}
H,GNN-homo,GNN-hete,DetMDN,BMDN-VI,BMDN-DE,BMDN-DEVI
0,0.1690,0.1689,0.1625,0.148,0.1565,0.1478
1,0.2077,0.2063,0.1947,0.1983,0.1907,0.189
2,0.3429,0.2882,0.2825,0.3025,0.2838,0.2937
3,0.3089,0.2617,0.2501,0.2359,0.2451,0.2363
4,0.2627,0.2312,0.2071,0.1986,0.2035,0.1928
5,0.2912,0.2511,0.2316,0.2262,0.2312,0.2229
6,0.2500,0.2224,0.2055,0.2002,0.2038,0.2017
7,0.0781,0.0758,0.0235,0.0038,0.0186,0.0034
8,0.0555,0.0548,0.0449,0.0333,0.0386,0.0326
9,0.1078,0.1096,0.1017,0.106,0.1009,0.1036
10,0.1535,0.1471,0.151,0.1486,0.146,0.1434
11,0.2408,0.2123,0.2189,0.2255,0.2214,0.2201
12,0.1523,0.1479,0.1357,0.1364,0.1335,0.1341
13,0.0495,0.0401,0.031,0.0297,0.0301,0.0293
14,0.0459,0.0371,0.0295,0.028,0.0292,0.0277
15,0.0492,0.0424,0.0419,0.0418,0.042,0.0418
16,0.0433,0.0347,0.0314,0.0301,0.0303,0.0301
17,0.0446,0.0343,0.0306,0.0313,0.0306,0.031
18,0.0391,0.0251,0.0197,0.0223,0.0193,0.0226
19,0.0571,0.0526,0.0523,0.0539,0.052,0.0521
20,0.0338,0.0157,0.0091,0.0138,0.0095,0.0124
21,0.0370,0.0234,0.0258,0.0245,0.0253,0.0226
22,0.0354,0.0118,0.0096,0.0099,0.0096,0.0098
23,0.0339,0.0038,0.0024,0.002,0.0021,0.0019
\end{filecontents*}
\pgfplotstabletypeset[
		col sep=comma,
		string type,
		every head row/.style={%
			before row={\hline
			},
			after row=\hline
		},
		every last row/.style={after row=\hline},
		columns/name/.style={column name=Name, column type=l},
		columns/surname/.style={column name=Surname, column type=l},
		columns/age/.style={column name=Age, column type=c},
		]{figures/tables/hourlyCRPS5.csv}
	}
	\label{crps_hour_uk5}
\end{table}

\begin{table}[h]
	\small
	\centering
	\caption{Hourly CRPS performance on UK-SMEC-H\#6 Test set.}
	\makebox[\linewidth]{
		\begin{filecontents*}{figures/tables/hourlyCRPS6.csv}
H,GNN-homo,GNN-hete,DetMDN,BMDN-VI,BMDN-DE,BMDN-DEVI
0,0.0439,0.0111,0.0115,0.0116,0.0117,0.0122
1,0.0448,0.0128,0.0131,0.013,0.0131,0.014
2,0.0449,0.0159,0.0155,0.0156,0.0154,0.0158
3,0.0442,0.012,0.0121,0.012,0.012,0.0126
4,0.0528,0.0186,0.018,0.0237,0.0176,0.0257
5,0.0604,0.0495,0.0406,0.0462,0.0389,0.0419
6,0.0459,0.0148,0.0142,0.0149,0.0144,0.0148
7,0.0673,0.0678,0.0691,0.0646,0.0645,0.0599
8,0.0771,0.0738,0.0716,0.0697,0.0678,0.0701
9,0.0645,0.0347,0.0273,0.0252,0.0271,0.0267
10,0.0931,0.0961,0.0939,0.0891,0.0903,0.0869
11,0.1169,0.1209,0.1157,0.1069,0.1141,0.1087
12,0.1442,0.1307,0.1338,0.1208,0.13,0.1227
13,0.1306,0.1236,0.1164,0.1144,0.1129,0.117
14,0.0739,0.0487,0.0492,0.0468,0.0508,0.0483
15,0.0534,0.0414,0.0424,0.0418,0.0427,0.0364
16,0.0866,0.0737,0.0642,0.0571,0.0603,0.0578
17,0.1333,0.1278,0.137,0.1255,0.1321,0.1171
18,0.1995,0.2263,0.2046,0.1969,0.1995,0.1917
19,0.4254,0.4369,0.3716,0.3516,0.3854,0.329
20,0.0937,0.074,0.0712,0.0574,0.0662,0.0664
21,0.0531,0.0242,0.0248,0.024,0.0235,0.025
22,0.0465,0.0205,0.0211,0.0205,0.0209,0.0201
23,0.0446,0.0178,0.0174,0.0176,0.0173,0.0176
\end{filecontents*}
\pgfplotstabletypeset[
		col sep=comma,
		string type,
		every head row/.style={%
			before row={\hline
			},
			after row=\hline
		},
		every last row/.style={after row=\hline},
		columns/name/.style={column name=Name, column type=l},
		columns/surname/.style={column name=Surname, column type=l},
		columns/age/.style={column name=Age, column type=c},
		]{figures/tables/hourlyCRPS6.csv}
	}
	\label{crps_hour_uk6}
\end{table}

\begin{table}[h]
	\small
	\centering
	\caption{Hourly CRPS performance on UK-SMEC-H\#7 Test set.}
	\makebox[\linewidth]{
		\begin{filecontents*}{figures/tables/hourlyCRPS7.csv}
H,GNN-homo,GNN-hete,DetMDN,BMDN-VI,BMDN-DE,BMDN-DEVI
0,0.1298,0.0875,0.0914,0.0696,0.0714,0.0616
1,0.3233,0.31,0.3047,0.2982,0.3002,0.2899
2,0.1443,0.1873,0.1943,0.1513,0.1677,0.1552
3,0.1873,0.1973,0.1947,0.1748,0.1864,0.1738
4,0.1729,0.163,0.1406,0.1371,0.1395,0.135
5,0.3327,0.302,0.287,0.2968,0.2936,0.2935
6,0.2936,0.2697,0.2653,0.2828,0.2697,0.276
7,0.0614,0.0521,0.0466,0.0384,0.0441,0.04
8,0.0556,0.0398,0.0406,0.039,0.04,0.0393
9,0.0554,0.0372,0.0363,0.0358,0.0369,0.0357
10,0.0572,0.038,0.0374,0.0369,0.0376,0.0357
11,0.0511,0.0295,0.0229,0.0232,0.0228,0.023
12,0.0575,0.0441,0.0328,0.0326,0.0327,0.0327
13,0.0923,0.0798,0.0788,0.0746,0.0789,0.0747
14,0.0702,0.0562,0.0533,0.0521,0.0543,0.0518
15,0.0581,0.0433,0.0421,0.0415,0.0427,0.0416
16,0.0552,0.0401,0.0365,0.0358,0.0375,0.0365
17,0.1169,0.114,0.1065,0.1016,0.1057,0.1014
18,0.1499,0.1447,0.1405,0.1326,0.138,0.1311
19,0.0612,0.0483,0.0412,0.0458,0.0437,0.0454
20,0.0500,0.0242,0.0244,0.0313,0.0278,0.029
21,0.0497,0.0231,0.0228,0.0272,0.0237,0.0264
22,0.0501,0.0205,0.0219,0.0235,0.0208,0.0227
23,0.0502,0.0212,0.0207,0.0207,0.0214,0.021
\end{filecontents*}
\pgfplotstabletypeset[
		col sep=comma,
		string type,
		every head row/.style={%
			before row={\hline
			},
			after row=\hline
		},
		every last row/.style={after row=\hline},
		columns/name/.style={column name=Name, column type=l},
		columns/surname/.style={column name=Surname, column type=l},
		columns/age/.style={column name=Age, column type=c},
		]{figures/tables/hourlyCRPS7.csv}
	}
	\label{crps_hour_uk7}
\end{table}

\begin{table}[h]
	\small
	\centering
	\caption{Hourly CRPS performance on UK-SMEC-H\#8 Test set.}
	\makebox[\linewidth]{
		\begin{filecontents*}{figures/tables/hourlyCRPS8.csv}
H,GNN-homo,GNN-hete,DetMDN,BMDN-VI,BMDN-DE,BMDN-DEVI
0,0.2033,0.1995,0.1977,0.2059,0.2075,0.1969
1,0.1397,0.1579,0.1484,0.1725,0.1613,0.145
2,0.1962,0.2034,0.2023,0.1938,0.2042,0.1763
3,0.3000,0.2692,0.2857,0.2465,0.2645,0.2381
4,0.3889,0.3275,0.3376,0.3165,0.3241,0.3122
5,0.1557,0.178,0.1138,0.1558,0.1167,0.1437
6,0.1362,0.1585,0.1094,0.1263,0.1101,0.1195
7,0.0623,0.075,0.0374,0.0365,0.0354,0.0337
8,0.0585,0.0466,0.0451,0.0425,0.0428,0.0424
9,0.0590,0.0466,0.0384,0.0354,0.0371,0.0353
10,0.0735,0.0797,0.052,0.052,0.0518,0.0517
11,0.1090,0.1181,0.0915,0.0925,0.0926,0.0915
12,0.1381,0.1402,0.1261,0.1226,0.1244,0.123
13,0.0815,0.0839,0.0684,0.069,0.0686,0.0675
14,0.0630,0.0589,0.0399,0.0399,0.0396,0.0397
15,0.0632,0.0581,0.0431,0.0426,0.0424,0.0426
16,0.0711,0.0618,0.0521,0.0521,0.0516,0.0518
17,0.0575,0.0413,0.042,0.0411,0.0418,0.041
18,0.0527,0.0293,0.0279,0.0278,0.0277,0.0278
19,0.0466,0.0178,0.0173,0.0175,0.0174,0.0173
20,0.0461,0.0166,0.0161,0.0165,0.016,0.016
21,0.0464,0.0202,0.0198,0.0203,0.02,0.02
22,0.0478,0.025,0.0245,0.0247,0.0247,0.0245
23,0.0493,0.0239,0.0205,0.0206,0.0205,0.0205
\end{filecontents*}
\pgfplotstabletypeset[
		col sep=comma,
		string type,
		every head row/.style={%
			before row={\hline
			},
			after row=\hline
		},
		every last row/.style={after row=\hline},
		columns/name/.style={column name=Name, column type=l},
		columns/surname/.style={column name=Surname, column type=l},
		columns/age/.style={column name=Age, column type=c},
		]{figures/tables/hourlyCRPS8.csv}
	}
	\label{crps_hour_uk8}
\end{table}
 
 \begin{table}[h]
 	\small
 	\centering
 	\caption{Daily CRPS performance on UK-SMEC-H\#1 Test set.}
 	\makebox[\linewidth]{
 		\begin{filecontents*}{figures/tables/dailyCRPS1.csv}
D,GNN-homo,GNN-hete,DetMDN,BMDN-VI,BMDN-DE,BMDN-DEVI
0,0.1512,0.1444,0.138,0.1394,0.1364,0.1365
1,0.1397,0.1326,0.1282,0.127,0.1246,0.1243
2,0.1489,0.1417,0.1371,0.1375,0.1341,0.1337
3,0.1444,0.1339,0.126,0.1272,0.124,0.1241
4,0.1324,0.1276,0.1219,0.1228,0.1189,0.1198
5,0.1624,0.1549,0.1475,0.1464,0.1455,0.1438
6,0.1606,0.1535,0.1509,0.1525,0.1489,0.1481
\end{filecontents*}
\pgfplotstabletypeset[
 		col sep=comma,
 		string type,
 		every head row/.style={%
 			before row={\hline
 			},
 			after row=\hline
 		},
 		every last row/.style={after row=\hline},
 		columns/name/.style={column name=Name, column type=l},
 		columns/surname/.style={column name=Surname, column type=l},
 		columns/age/.style={column name=Age, column type=c},
 		]{figures/tables/dailyCRPS1.csv}
 	}
 	\label{crps_day_uk1}
 \end{table}

 \begin{table}[h]
	\small
	\centering
	\caption{Daily CRPS performance on UK-SMEC-H\#2 Test set.}
	\makebox[\linewidth]{
		\begin{filecontents*}{figures/tables/dailyCRPS2.csv}
D,GNN-homo,GNN-hete,DetMDN,BMDN-VI,BMDN-DE,BMDN-DEVI
0,0.2306,0.2095,0.1986,0.1987,0.1985,0.1974
1,0.208,0.1925,0.185,0.1827,0.1841,0.182
2,0.2326,0.2074,0.1997,0.1964,0.1986,0.1957
3,0.2163,0.1978,0.1869,0.1883,0.1861,0.1873
4,0.225,0.2059,0.1967,0.1952,0.1959,0.1952
5,0.2347,0.2148,0.2081,0.2042,0.206,0.2044
6,0.2365,0.2106,0.2014,0.1993,0.1998,0.1989
\end{filecontents*}
\pgfplotstabletypeset[
		col sep=comma,
		string type,
		every head row/.style={%
			before row={\hline
			},
			after row=\hline
		},
		every last row/.style={after row=\hline},
		columns/name/.style={column name=Name, column type=l},
		columns/surname/.style={column name=Surname, column type=l},
		columns/age/.style={column name=Age, column type=c},
		]{figures/tables/dailyCRPS2.csv}
	}
	\label{crps_day_uk2}
\end{table}

 \begin{table}[h]
	\small
	\centering
	\caption{Daily CRPS performance on UK-SMEC-H\#3 Test set.}
	\makebox[\linewidth]{
		\begin{filecontents*}{figures/tables/dailyCRPS3.csv}
D,GNN-homo,GNN-hete,DetMDN,BMDN-VI,BMDN-DE,BMDN-DEVI
0,0.173,0.1555,0.1455,0.1404,0.1452,0.1408
1,0.1654,0.1518,0.1456,0.1412,0.1425,0.1419
2,0.1587,0.1436,0.1308,0.1319,0.1325,0.1312
3,0.1482,0.1363,0.1273,0.1259,0.1257,0.124
4,0.1722,0.1584,0.1561,0.1528,0.1542,0.1504
5,0.1706,0.1547,0.1424,0.1413,0.1427,0.1389
6,0.1914,0.1692,0.1597,0.1565,0.1599,0.1579
\end{filecontents*}
\pgfplotstabletypeset[
		col sep=comma,
		string type,
		every head row/.style={%
			before row={\hline
			},
			after row=\hline
		},
		every last row/.style={after row=\hline},
		columns/name/.style={column name=Name, column type=l},
		columns/surname/.style={column name=Surname, column type=l},
		columns/age/.style={column name=Age, column type=c},
		]{figures/tables/dailyCRPS3.csv}
	}
	\label{crps_day_uk3}
\end{table}

 \begin{table}[h]
	\small
	\centering
	\caption{Daily CRPS performance on UK-SMEC-H\#4 Test set.}
	\makebox[\linewidth]{
		\begin{filecontents*}{figures/tables/dailyCRPS4.csv}
D,GNN-homo,GNN-hete,DetMDN,BMDN-VI,BMDN-DE,BMDN-DEVI
0,0.1013,0.0826,0.0762,0.0765,0.0778,0.0742
1,0.1005,0.0927,0.0835,0.0796,0.0818,0.0781
2,0.093,0.0845,0.0717,0.0734,0.0705,0.0707
3,0.092,0.0797,0.067,0.0688,0.0672,0.067
4,0.1003,0.0812,0.0746,0.0735,0.0753,0.072
5,0.0907,0.0798,0.0682,0.0676,0.0667,0.0666
6,0.101,0.0918,0.0862,0.0828,0.0841,0.0798
\end{filecontents*}
\pgfplotstabletypeset[
		col sep=comma,
		string type,
		every head row/.style={%
			before row={\hline
			},
			after row=\hline
		},
		every last row/.style={after row=\hline},
		columns/name/.style={column name=Name, column type=l},
		columns/surname/.style={column name=Surname, column type=l},
		columns/age/.style={column name=Age, column type=c},
		]{figures/tables/dailyCRPS4.csv}
	}
	\label{crps_day_uk4}
\end{table}

 \begin{table}[h]
	\small
	\centering
	\caption{Daily CRPS performance on UK-SMEC-H\#5 Test set.}
	\makebox[\linewidth]{
		\begin{filecontents*}{figures/tables/dailyCRPS5.csv}
D,GNN-homo,GNN-hete,DetMDN,BMDN-VI,BMDN-DE,BMDN-DEVI
0,0.123,0.1093,0.0986,0.0958,0.0969,0.0948
1,0.1312,0.1143,0.1067,0.1043,0.1041,0.1017
2,0.1251,0.1093,0.1016,0.1029,0.1001,0.1006
3,0.1282,0.11,0.1014,0.099,0.0992,0.0975
4,0.1296,0.1114,0.1031,0.103,0.1023,0.1018
5,0.1239,0.1079,0.1028,0.1011,0.1004,0.0985
6,0.1387,0.1232,0.1116,0.1073,0.1111,0.1048
\end{filecontents*}
\pgfplotstabletypeset[
		col sep=comma,
		string type,
		every head row/.style={%
			before row={\hline
			},
			after row=\hline
		},
		every last row/.style={after row=\hline},
		columns/name/.style={column name=Name, column type=l},
		columns/surname/.style={column name=Surname, column type=l},
		columns/age/.style={column name=Age, column type=c},
		]{figures/tables/dailyCRPS5.csv}
	}
	\label{crps_day_uk5}
\end{table}

 \begin{table}[h]
	\small
	\centering
	\caption{Daily CRPS performance on UK-SMEC-H\#6 Test set.}
	\makebox[\linewidth]{
		\begin{filecontents*}{figures/tables/dailyCRPS6.csv}
D,GNN-homo,GNN-hete,DetMDN,BMDN-VI,BMDN-DE,BMDN-DEVI
0,0.1028,0.0904,0.0824,0.0801,0.0837,0.0735
1,0.0755,0.0631,0.0502,0.0449,0.0489,0.0453
2,0.1165,0.1087,0.11,0.1009,0.1062,0.1024
3,0.0726,0.0515,0.0501,0.0459,0.0465,0.0478
4,0.0734,0.0664,0.0552,0.0607,0.058,0.0568
5,0.1104,0.0893,0.0881,0.0779,0.0888,0.0789
6,0.1028,0.0784,0.0767,0.0771,0.0742,0.0743
\end{filecontents*}
\pgfplotstabletypeset[
		col sep=comma,
		string type,
		every head row/.style={%
			before row={\hline
			},
			after row=\hline
		},
		every last row/.style={after row=\hline},
		columns/name/.style={column name=Name, column type=l},
		columns/surname/.style={column name=Surname, column type=l},
		columns/age/.style={column name=Age, column type=c},
		]{figures/tables/dailyCRPS6.csv}
	}
	\label{crps_day_uk6}
\end{table}

 \begin{table}[h]
	\small
	\centering
	\caption{Daily CRPS performance on UK-SMEC-H\#7 Test set.}
	\makebox[\linewidth]{
		\begin{filecontents*}{figures/tables/dailyCRPS7.csv}
D,GNN-homo,GNN-hete,DetMDN,BMDN-VI,BMDN-DE,BMDN-DEVI
0,0.1199,0.1002,0.0975,0.098,0.0978,0.0942
1,0.1055,0.092,0.088,0.0824,0.0853,0.0823
2,0.1214,0.1072,0.1034,0.1017,0.1015,0.1013
3,0.1182,0.1023,0.1006,0.0947,0.0982,0.0932
4,0.1107,0.0959,0.0911,0.0902,0.0885,0.0881
5,0.1044,0.0938,0.0882,0.0845,0.0855,0.0836
6,0.1148,0.1005,0.0968,0.0908,0.0954,0.0909
\end{filecontents*}
\pgfplotstabletypeset[
		col sep=comma,
		string type,
		every head row/.style={%
			before row={\hline
			},
			after row=\hline
		},
		every last row/.style={after row=\hline},
		columns/name/.style={column name=Name, column type=l},
		columns/surname/.style={column name=Surname, column type=l},
		columns/age/.style={column name=Age, column type=c},
		]{figures/tables/dailyCRPS7.csv}
	}
	\label{crps_day_uk7}
\end{table}

 \begin{table}[h]
	\small
	\centering
	\caption{Daily CRPS performance on UK-SMEC-H\#8 Test set.}
	\makebox[\linewidth]{
		\begin{filecontents*}{figures/tables/dailyCRPS8.csv}
D,GNN-homo,GNN-hete,DetMDN,BMDN-VI,BMDN-DE,BMDN-DEVI
0,0.0952,0.0931,0.0773,0.0826,0.0778,0.0777
1,0.1103,0.1024,0.0902,0.0923,0.0896,0.087
2,0.1094,0.0997,0.0913,0.0864,0.0893,0.0833
3,0.1189,0.1098,0.0948,0.0998,0.0939,0.0943
4,0.1236,0.1095,0.1033,0.0962,0.0997,0.0949
5,0.1126,0.101,0.0915,0.0911,0.0936,0.0888
6,0.0993,0.0932,0.0786,0.0827,0.0792,0.0781
\end{filecontents*}
\pgfplotstabletypeset[
		col sep=comma,
		string type,
		every head row/.style={%
			before row={\hline
			},
			after row=\hline
		},
		every last row/.style={after row=\hline},
		columns/name/.style={column name=Name, column type=l},
		columns/surname/.style={column name=Surname, column type=l},
		columns/age/.style={column name=Age, column type=c},
		]{figures/tables/dailyCRPS8.csv}
	}
	\label{crps_day_uk8}
\end{table}

\section{Conclusions and next developments}
In this paper we have presented a novel approach to probabilistic load forecasting (PLF) based on Bayesian deep learning techniques, capturing both aleatoric and epistemic uncertainty contributions within the model predictions. The inherent stochasticity of the electric load has been addressed by a full conditional density estimation, providing input features dependent representations. To this end, we deployed a flexible Mixture Density Network architecture, including spherical Gaussian kernels and a proper configuration of the last hidden layer, to guarantee both positive definite variances and valid categorical distributions for components mixing. Then, point estimation in the parameters space, given by conventional maximum likelihood training approaches, has been extended into posterior distributions inference through a Bayesian framework. Hence, the weights are intrinsically considered as stochastic variables, marginalized within the function space distribution during prediction, thus conveying model confidence from the features space up to the network predictions. Hence, both a principled approach to epistemic uncertainty integration as well as an intrinsic regularization effect have been obtained, resulting particularly crucial when complex neural network models are adopted for PLF. 
 
Since standard techniques feasible for simple Bayesian regression models and small data regimes are not computationally scalable for deep learning applications, we leveraged on Variational inference. Then, the Bayesian MDN inference tasks is tackled through an end-to-end training procedure, minimizing the Evidence Lowe Bound (ELBO) with reference to the variational parameters of a Mean Field approximation. Thus, function space representation capabilities, cheaper computational costs and mapping flexibility are concurrently addressed, fundamental to properly estimate the articulated relations between the conditioning features and the target load distribution. Besides, efficient parameters optimization via standard back-propagation routines is enabled, by exploiting the re-parametrization trick. To avoid the potential mis-specification of complex neural networks in finite samples conditions, we incorporated a tempered posterior in the inference process, leading to a weighted ELBO optimization. Deep neural networks ensembles have been considered to improve posterior marginalization, by covering samples from different modes,  exploiting parallel model training procedures, starting from different random initialization and data shuffles. Then, we introduced an integrated approach based on a Mean Field-Bayesian MDN ensemble, to achieve both local and global approximation capabilities within a structured inference machinery.

We evaluated the proposed PLF approach on publicly available case studies, targeting short term forecasting at fine-grained single households consumption scale. A detailed statistical analysis of the considered data setting has been performed, since lacking in the available literature, to extract the major characteristics of the overall distributions, support model configuration and explanation of the results. 
Application scenarios have been framed in day-ahead prediction tasks over the next 24 hours, adopting CRPS to achieve proper scoring of the experimental results, integrating both sharpness and calibration requirements. We demonstrated the capability of proposed approach to achieve robust performances in out-of-sample conditions, reporting detailed quantitative evaluation of different model settings as well as comparison to state of the art PLF techniques.

Actually, we envision this paper as a first step towards the full exploration of Bayesian Mixture Density Networks for probabilistic load forecast. In fact, various future extensions are foreseen, here briefly summarized. In particular, we plan to investigate alternative network architectures, different kernels form in probabilistic layers as well as further inference techniques, exploiting different priors and posterior approximations. The integration of more specific conditioning variables and hyperparameters configurations is key to further improve prediction performance in each application case, which would require the implementation of advanced search algorithms for efficient space exploration. Novel techniques to foster diversity in the ensembles, improved posterior modes coverage and function space marginalization are interesting directions to be explored as well, considering also data augmentation and different sub-models combinations. Moreover, we foresee the application to further probabilistic forecasting problems.

\section*{References}
\bibliography{mybibfile_v01}

\begin{thebibliography}{10}
\expandafter\ifx\csname url\endcsname\relax
  \def\url#1{\texttt{#1}}\fi
\expandafter\ifx\csname urlprefix\endcsname\relax\def\urlprefix{URL }\fi
\expandafter\ifx\csname href\endcsname\relax
  \def\href#1#2{#2} \def\path#1{#1}\fi

\bibitem{HE2020114396}
F.~He, J.~Zhou, L.~Mo, K.~Feng, G.~Liu, Z.~He,
  \href{http://www.sciencedirect.com/science/article/pii/S0306261919320835}{Day-ahead
  short-term load probability density forecasting method with a
  decomposition-based quantile regression forest}, Applied Energy 262 (2020)
  114396.
\newblock \href
  {http://dx.doi.org/https://doi.org/10.1016/j.apenergy.2019.114396}
  {\path{doi:https://doi.org/10.1016/j.apenergy.2019.114396}}.
\newline\urlprefix\url{http://www.sciencedirect.com/science/article/pii/S0306261919320835}

\bibitem{hong2016probabilistic}
T.~Hong, S.~Fan, Probabilistic electric load forecasting: A tutorial review,
  International Journal of Forecasting 32~(3) (2016) 914--938.

\bibitem{weronbook}
R.~Weron, \href{https://ideas.repec.org/b/wuu/hsbook/hsbook0601.html}{{Modeling
  and Forecasting Electricity Loads and Prices: A Statistical Approach}}, no.
  hsbook0601 in HSC Books, Hugo Steinhaus Center, Wroclaw University of
  Technology, 2006.
\newline\urlprefix\url{https://ideas.repec.org/b/wuu/hsbook/hsbook0601.html}

\bibitem{expoSmoth}
R.~Göb, K.~Lurz, A.~Pievatolo, Electrical load forecasting by exponential
  smoothing with covariates, Applied Stochastic Models in Business and Industry
  29~(6) (2013) 629--645.
\newblock \href {http://dx.doi.org/https://doi.org/10.1002/asmb.2008}
  {\path{doi:https://doi.org/10.1002/asmb.2008}}.

\bibitem{GAILLARD20161038}
P.~Gaillard, Y.~Goude, R.~Nedellec,
  \href{http://www.sciencedirect.com/science/article/pii/S0169207015001545}{Additive
  models and robust aggregation for gefcom2014 probabilistic electric load and
  electricity price forecasting}, International Journal of Forecasting 32~(3)
  (2016) 1038 -- 1050.
\newblock \href
  {http://dx.doi.org/https://doi.org/10.1016/j.ijforecast.2015.12.001}
  {\path{doi:https://doi.org/10.1016/j.ijforecast.2015.12.001}}.
\newline\urlprefix\url{http://www.sciencedirect.com/science/article/pii/S0169207015001545}

\bibitem{YANG2018499}
Y.~Yang, S.~Li, W.~Li, M.~Qu,
  \href{http://www.sciencedirect.com/science/article/pii/S0306261917316100}{Power
  load probability density forecasting using gaussian process quantile
  regression}, Applied Energy 213 (2018) 499 -- 509.
\newblock \href
  {http://dx.doi.org/https://doi.org/10.1016/j.apenergy.2017.11.035}
  {\path{doi:https://doi.org/10.1016/j.apenergy.2017.11.035}}.
\newline\urlprefix\url{http://www.sciencedirect.com/science/article/pii/S0306261917316100}

\bibitem{BENTAIEB2014382}
S.~{Ben Taieb}, R.~J. Hyndman,
  \href{http://www.sciencedirect.com/science/article/pii/S0169207013000812}{A
  gradient boosting approach to the kaggle load forecasting competition},
  International Journal of Forecasting 30~(2) (2014) 382 -- 394.
\newblock \href
  {http://dx.doi.org/https://doi.org/10.1016/j.ijforecast.2013.07.005}
  {\path{doi:https://doi.org/10.1016/j.ijforecast.2013.07.005}}.
\newline\urlprefix\url{http://www.sciencedirect.com/science/article/pii/S0169207013000812}

\bibitem{1350819}
{Bo-Juen Chen}, {Ming-Wei Chang}, {Chih-Jen lin}, Load forecasting using
  support vector machines: a study on eunite competition 2001, IEEE
  Transactions on Power Systems 19~(4) (2004) 1821--1830.
\newblock \href {http://dx.doi.org/10.1109/TPWRS.2004.835679}
  {\path{doi:10.1109/TPWRS.2004.835679}}.

\bibitem{LAHOUAR20151040}
A.~Lahouar, J.~{Ben Hadj Slama},
  \href{http://www.sciencedirect.com/science/article/pii/S0196890415006925}{Day-ahead
  load forecast using random forest and expert input selection}, Energy
  Conversion and Management 103 (2015) 1040 -- 1051.
\newblock \href
  {http://dx.doi.org/https://doi.org/10.1016/j.enconman.2015.07.041}
  {\path{doi:https://doi.org/10.1016/j.enconman.2015.07.041}}.
\newline\urlprefix\url{http://www.sciencedirect.com/science/article/pii/S0196890415006925}

\bibitem{5685606}
M.~{Rejc}, M.~{Pantos}, Short-term transmission-loss forecast for the slovenian
  transmission power system based on a fuzzy-logic decision approach, IEEE
  Transactions on Power Systems 26~(3) (2011) 1511--1521.
\newblock \href {http://dx.doi.org/10.1109/TPWRS.2010.2096829}
  {\path{doi:10.1109/TPWRS.2010.2096829}}.

\bibitem{8372953}
K.~{Chen}, K.~{Chen}, Q.~{Wang}, Z.~{He}, J.~{Hu}, J.~{He}, Short-term load
  forecasting with deep residual networks, IEEE Transactions on Smart Grid
  10~(4) (2019) 3943--3952.
\newblock \href {http://dx.doi.org/10.1109/TSG.2018.2844307}
  {\path{doi:10.1109/TSG.2018.2844307}}.

\bibitem{7124541}
S.~{Li}, P.~{Wang}, L.~{Goel}, A novel wavelet-based ensemble method for
  short-term load forecasting with hybrid neural networks and feature
  selection, IEEE Transactions on Power Systems 31~(3) (2016) 1788--1798.
\newblock \href {http://dx.doi.org/10.1109/TPWRS.2015.2438322}
  {\path{doi:10.1109/TPWRS.2015.2438322}}.

\bibitem{8322199}
Y.~{Wang}, Q.~{Chen}, T.~{Hong}, C.~{Kang}, Review of smart meter data
  analytics: Applications, methodologies, and challenges, IEEE Transactions on
  Smart Grid 10~(3) (2019) 3125--3148.
\newblock \href {http://dx.doi.org/10.1109/TSG.2018.2818167}
  {\path{doi:10.1109/TSG.2018.2818167}}.

\bibitem{HAHN2009902}
H.~Hahn, S.~Meyer-Nieberg, S.~Pickl,
  \href{http://www.sciencedirect.com/science/article/pii/S0377221709002094}{Electric
  load forecasting methods: Tools for decision making}, European Journal of
  Operational Research 199~(3) (2009) 902 -- 907.
\newblock \href {http://dx.doi.org/https://doi.org/10.1016/j.ejor.2009.01.062}
  {\path{doi:https://doi.org/10.1016/j.ejor.2009.01.062}}.
\newline\urlprefix\url{http://www.sciencedirect.com/science/article/pii/S0377221709002094}

\bibitem{BRUSAFERRI20191158}
A.~Brusaferri, M.~Matteucci, P.~Portolani, A.~Vitali,
  \href{http://www.sciencedirect.com/science/article/pii/S0306261919309237}{Bayesian
  deep learning based method for probabilistic forecast of day-ahead
  electricity prices}, Applied Energy 250 (2019) 1158 -- 1175.
\newblock \href
  {http://dx.doi.org/https://doi.org/10.1016/j.apenergy.2019.05.068}
  {\path{doi:https://doi.org/10.1016/j.apenergy.2019.05.068}}.
\newline\urlprefix\url{http://www.sciencedirect.com/science/article/pii/S0306261919309237}

\bibitem{8988175}
M.~{Afrasiabi}, M.~{Mohammadi}, M.~{Rastegar}, L.~{Stankovic}, S.~{Afrasiabi},
  M.~{Khazaei}, Deep-based conditional probability density function forecasting
  of residential loads, IEEE Transactions on Smart Grid 11~(4) (2020)
  3646--3657.
\newblock \href {http://dx.doi.org/10.1109/TSG.2020.2972513}
  {\path{doi:10.1109/TSG.2020.2972513}}.

\bibitem{MUNKHAMMAR2021116180}
J.~Munkhammar, D.~{van der Meer}, J.~Widén,
  \href{http://www.sciencedirect.com/science/article/pii/S0306261920315816}{Very
  short term load forecasting of residential electricity consumption using the
  markov-chain mixture distribution (mcm) model}, Applied Energy 282 (2021)
  116180.
\newblock \href
  {http://dx.doi.org/https://doi.org/10.1016/j.apenergy.2020.116180}
  {\path{doi:https://doi.org/10.1016/j.apenergy.2020.116180}}.
\newline\urlprefix\url{http://www.sciencedirect.com/science/article/pii/S0306261920315816}

\bibitem{RAMIN2018622}
D.~Ramin, S.~Spinelli, A.~Brusaferri,
  \href{http://www.sciencedirect.com/science/article/pii/S0306261918304227}{Demand-side
  management via optimal production scheduling in power-intensive industries:
  The case of metal casting process}, Applied Energy 225 (2018) 622 -- 636.
\newblock \href
  {http://dx.doi.org/https://doi.org/10.1016/j.apenergy.2018.03.084}
  {\path{doi:https://doi.org/10.1016/j.apenergy.2018.03.084}}.
\newline\urlprefix\url{http://www.sciencedirect.com/science/article/pii/S0306261918304227}

\bibitem{pmlr-v80-kuleshov18a}
V.~Kuleshov, N.~Fenner, S.~Ermon,
  \href{http://proceedings.mlr.press/v80/kuleshov18a.html}{Accurate
  uncertainties for deep learning using calibrated regression}, in: J.~Dy,
  A.~Krause (Eds.), Proceedings of the 35th International Conference on Machine
  Learning, Vol.~80 of Proceedings of Machine Learning Research, PMLR,
  Stockholmsmässan, Stockholm Sweden, 2018, pp. 2796--2804.
\newline\urlprefix\url{http://proceedings.mlr.press/v80/kuleshov18a.html}

\bibitem{guo}
C.~Guo, G.~Pleiss, Y.~Sun, K.~Q. Weinberger, On calibration of modern neural
  networks, in: Proceedings of the 34th International Conference on Machine
  Learning - Volume 70, ICML'17, JMLR.org, 2017, p. 1321–1330.

\bibitem{NEURIPS2019_8558cb40}
Y.~Ovadia, E.~Fertig, J.~Ren, Z.~Nado, D.~Sculley, S.~Nowozin, J.~Dillon,
  B.~Lakshminarayanan, J.~Snoek, Can you trust your model's uncertainty?
  evaluating predictive uncertainty under dataset shift, in: Advances in Neural
  Information Processing Systems, Vol.~32, Curran Associates, Inc., 2019, pp.
  13991--14002.

\bibitem{gal1}
A.~Kendall, Y.~Gal, What uncertainties do we need in bayesian deep learning for
  computer vision?, in: Proceedings of the 31st International Conference on
  Neural Information Processing Systems, NIPS'17, Curran Associates Inc., Red
  Hook, NY, USA, 2017, p. 5580–5590.

\bibitem{wilson}
A.~G. Wilson, P.~Izmailov, Bayesian deep learning and a probabilistic
  perspective of generalization, in: Advances in Neural Information Processing
  Systems, Curran Associates, Inc., 2020.

\bibitem{wang2019probabilistic}
Y.~Wang, D.~Gan, M.~Sun, N.~Zhang, Z.~Lu, C.~Kang, Probabilistic individual
  load forecasting using pinball loss guided lstm, Applied Energy 235 (2019)
  10--20.

\bibitem{8810811}
A.~{Elvers}, M.~{Voß}, S.~{Albayrak}, Short-term probabilistic load
  forecasting at low aggregation levels using convolutional neural networks,
  in: 2019 IEEE Milan PowerTech, 2019, pp. 1--6.
\newblock \href {http://dx.doi.org/10.1109/PTC.2019.8810811}
  {\path{doi:10.1109/PTC.2019.8810811}}.

\bibitem{9024710}
D.~{Gan}, Y.~{Wang}, S.~{Yang}, C.~{Kang}, Embedding based quantile regression
  neural network for probabilistic load forecasting, Journal of Modern Power
  Systems and Clean Energy 6~(2) (2018) 244--254.
\newblock \href {http://dx.doi.org/10.1007/s40565-018-0380-x}
  {\path{doi:10.1007/s40565-018-0380-x}}.

\bibitem{zhang2018improved}
W.~Zhang, H.~Quan, D.~Srinivasan, An improved quantile regression neural
  network for probabilistic load forecasting, IEEE Transactions on Smart Grid
  10~(4) (2018) 4425--4434.

\bibitem{yang2019deep}
Y.~Yang, W.~Hong, S.~Li, Deep ensemble learning based probabilistic load
  forecasting in smart grids, Energy 189 (2019) 116324.

\bibitem{afrasiabi2020deep}
M.~Afrasiabi, M.~Mohammadi, M.~Rastegar, L.~Stankovic, S.~Afrasiabi,
  M.~Khazaei, Deep-based conditional probability density function forecasting
  of residential loads, IEEE Transactions on Smart Grid.

\bibitem{guo2018deep}
Z.~Guo, K.~Zhou, X.~Zhang, S.~Yang, A deep learning model for short-term power
  load and probability density forecasting, Energy 160 (2018) 1186--1200.

\bibitem{he2019electricity}
Y.~He, Y.~Qin, S.~Wang, X.~Wang, C.~Wang, Electricity consumption probability
  density forecasting method based on lasso-quantile regression neural network,
  Applied energy 233 (2019) 565--575.

\bibitem{zhang2020load}
S.~Zhang, Y.~Wang, Y.~Zhang, D.~Wang, N.~Zhang, Load probability density
  forecasting by transforming and combining quantile forecasts, Applied Energy
  277 (2020) 115600.

\bibitem{li2019combining}
T.~Li, Y.~Wang, N.~Zhang, Combining probability density forecasts for power
  electrical loads, IEEE Transactions on Smart Grid 11~(2) (2019) 1679--1690.

\bibitem{Bishop:2006:PRM:1162264}
C.~M. Bishop, Pattern Recognition and Machine Learning (Information Science and
  Statistics), Springer-Verlag, Berlin, Heidelberg, 2006.

\bibitem{liberty}
S.~Farquhar, L.~Smith, Y.~Gal, in: Advances in Neural Information Processing
  Systems, Curran Associates, Inc.

\bibitem{pmlr-v115-izmailov20a}
P.~Izmailov, W.~J. Maddox, P.~Kirichenko, T.~Garipov, D.~Vetrov, A.~G. Wilson,
  \href{http://proceedings.mlr.press/v115/izmailov20a.html}{Subspace inference
  for bayesian deep learning}, in: R.~P. Adams, V.~Gogate (Eds.), Proceedings
  of The 35th Uncertainty in Artificial Intelligence Conference, Vol. 115 of
  Proceedings of Machine Learning Research, PMLR, Tel Aviv, Israel, 2020, pp.
  1169--1179.
\newline\urlprefix\url{http://proceedings.mlr.press/v115/izmailov20a.html}

\bibitem{10.1145/3409383}
H.~Wang, D.-Y. Yeung, \href{https://doi.org/10.1145/3409383}{A survey on
  bayesian deep learning}, ACM Comput. Surv. 53~(5).
\newblock \href {http://dx.doi.org/10.1145/3409383}
  {\path{doi:10.1145/3409383}}.
\newline\urlprefix\url{https://doi.org/10.1145/3409383}

\bibitem{8743433}
M.~{Sun}, T.~{Zhang}, Y.~{Wang}, G.~{Strbac}, C.~{Kang}, Using bayesian deep
  learning to capture uncertainty for residential net load forecasting, IEEE
  Transactions on Power Systems 35~(1) (2020) 188--201.
\newblock \href {http://dx.doi.org/10.1109/TPWRS.2019.2924294}
  {\path{doi:10.1109/TPWRS.2019.2924294}}.

\bibitem{8462978}
S.~{Choi}, K.~{Lee}, S.~{Lim}, S.~{Oh}, Uncertainty-aware learning from
  demonstration using mixture density networks with sampling-free variance
  modeling, in: 2018 IEEE International Conference on Robotics and Automation
  (ICRA), 2018, pp. 6915--6922.
\newblock \href {http://dx.doi.org/10.1109/ICRA.2018.8462978}
  {\path{doi:10.1109/ICRA.2018.8462978}}.

\bibitem{Goodfellow-et-al-2016}
I.~Goodfellow, Y.~Bengio, A.~Courville, Deep Learning, MIT Press, 2016.

\bibitem{bishop94}
C.~M. Bishop,
  \href{https://publications.aston.ac.uk/id/eprint/373/1/NCRG_94_004.pdf}{Mixture
  density networks}, Research report, Aston University, Neural Computing
  Research Group (1994).
\newline\urlprefix\url{https://publications.aston.ac.uk/id/eprint/373/1/NCRG_94_004.pdf}

\bibitem{Makansi_2019_CVPR}
O.~Makansi, E.~Ilg, O.~Cicek, T.~Brox, Overcoming limitations of mixture
  density networks: A sampling and fitting framework for multimodal future
  prediction, in: Proceedings of the IEEE/CVF Conference on Computer Vision and
  Pattern Recognition (CVPR), 2019.

\bibitem{9150658}
F.~K. {Gustafsson}, M.~{Danelljan}, T.~B. {Schon}, Evaluating scalable bayesian
  deep learning methods for robust computer vision, in: 2020 IEEE/CVF
  Conference on Computer Vision and Pattern Recognition Workshops (CVPRW),
  2020, pp. 1289--1298.
\newblock \href {http://dx.doi.org/10.1109/CVPRW50498.2020.00167}
  {\path{doi:10.1109/CVPRW50498.2020.00167}}.

\bibitem{Neal:1996:BLN:525544}
R.~M. Neal, Bayesian Learning for Neural Networks, Springer-Verlag, Berlin,
  Heidelberg, 1996.

\bibitem{minka}
T.~Minka, Bayesian model averaging is not model combination.

\bibitem{817982}
L.~U. {Hjorth}, I.~T. {Nabney}, Regularisation of mixture density networks, in:
  1999 Ninth International Conference on Artificial Neural Networks ICANN 99.
  (Conf. Publ. No. 470), 1999, pp. 521--526 vol.2.
\newblock \href {http://dx.doi.org/10.1049/cp:19991162}
  {\path{doi:10.1049/cp:19991162}}.

\bibitem{journals/corr/Graves13}
A.~Graves,
  \href{http://dblp.uni-trier.de/db/journals/corr/corr1308.html#Graves13}{Generating
  sequences with recurrent neural networks.}, CoRR abs/1308.0850.
\newline\urlprefix\url{http://dblp.uni-trier.de/db/journals/corr/corr1308.html#Graves13}

\bibitem{2018arXiv180701987G}
S.~Fruhwirth-Schnatter, G.~Celeux, C.~P. Robert, Handbook of Mixture Analysis,
  Chapman and Hall/CRC, 2019.

\bibitem{10.1007/978-3-319-77583-8_11}
C.~P. Martin, J.~Torresen, Robojam: A musical mixture density network for
  collaborative touchscreen interaction, in: A.~Liapis, J.~J. Romero~Cardalda,
  A.~Ek{\'a}rt (Eds.), Computational Intelligence in Music, Sound, Art and
  Design, Springer International Publishing, Cham, 2018, pp. 161--176.

\bibitem{quality_BDL}
J.~Yao, W.~Pan, S.~Ghosh, F.~Doshi-Velez, Quality of uncertainty quantification
  for bayesian neural network inference, Thirty-sixth International Conference
  on Machine Learning, 2019, Workshop on Uncertainty and Robustness in Deep
  Learning.

\bibitem{10.5555/3104482.3104568}
M.~Welling, Y.~W. Teh, Bayesian learning via stochastic gradient langevin
  dynamics, in: Proceedings of the 28th International Conference on
  International Conference on Machine Learning, ICML'11, Omnipress, Madison,
  WI, USA, 2011, p. 681–688.

\bibitem{pmlr-v32-cheni14}
T.~Chen, E.~Fox, C.~Guestrin, Stochastic gradient hamiltonian monte carlo, in:
  E.~P. Xing, T.~Jebara (Eds.), Proceedings of the 31st International
  Conference on Machine Learning, Vol.~32 of Proceedings of Machine Learning
  Research, PMLR, Bejing, China, 2014, pp. 1683--1691.

\bibitem{10.1145/168304.168306}
G.~E. Hinton, D.~van Camp, \href{https://doi.org/10.1145/168304.168306}{Keeping
  the neural networks simple by minimizing the description length of the
  weights}, in: Proceedings of the Sixth Annual Conference on Computational
  Learning Theory, COLT '93, Association for Computing Machinery, New York, NY,
  USA, 1993, p. 5–13.
\newblock \href {http://dx.doi.org/10.1145/168304.168306}
  {\path{doi:10.1145/168304.168306}}.
\newline\urlprefix\url{https://doi.org/10.1145/168304.168306}

\bibitem{pmlr-v108-farquhar20a}
S.~Farquhar, M.~A. Osborne, Y.~Gal, Radial bayesian neural networks: Beyond
  discrete support in large-scale bayesian deep learning, in: S.~Chiappa,
  R.~Calandra (Eds.), Proceedings of the Twenty Third International Conference
  on Artificial Intelligence and Statistics, Vol. 108 of Proceedings of Machine
  Learning Research, PMLR, Online, 2020, pp. 1352--1362.

\bibitem{pmlr-v115-hafner20a}
D.~Hafner, D.~Tran, T.~Lillicrap, A.~Irpan, J.~Davidson,
  \href{http://proceedings.mlr.press/v115/hafner20a.html}{Noise contrastive
  priors for functional uncertainty}, in: R.~P. Adams, V.~Gogate (Eds.),
  Proceedings of The 35th Uncertainty in Artificial Intelligence Conference,
  Vol. 115 of Proceedings of Machine Learning Research, PMLR, Tel Aviv, Israel,
  2020, pp. 905--914.
\newline\urlprefix\url{http://proceedings.mlr.press/v115/hafner20a.html}

\bibitem{Blundell:2015:WUN:3045118.3045290}
C.~Blundell, J.~Cornebise, K.~Kavukcuoglu, D.~Wierstra,
  \href{http://dl.acm.org/citation.cfm?id=3045118.3045290}{Weight uncertainty
  in neural networks}, in: 32Nd Int. Conf. on Machine Learning, ICML'15, 2015,
  pp. 1613--1622.
\newline\urlprefix\url{http://dl.acm.org/citation.cfm?id=3045118.3045290}

\bibitem{wen2018flipout}
Y.~Wen, P.~Vicol, J.~Ba, D.~Tran, R.~Grosse,
  \href{https://arxiv.org/abs/1803.04386}{Flipout: Efficient pseudo-independent
  weight perturbations on mini-batches}, in: International Conference on
  Learning Representations (ICLR), 2018.
\newblock \href {http://arxiv.org/abs/1803.04386} {\path{arXiv:1803.04386}}.
\newline\urlprefix\url{https://arxiv.org/abs/1803.04386}

\bibitem{wenzel20a}
F.~Wenzel, K.~Roth, B.~Veeling, J.~Swiatkowski, L.~Tran, S.~Mandt, J.~Snoek,
  T.~Salimans, R.~Jenatton, S.~Nowozin, How good is the {B}ayes posterior in
  deep neural networks really?, in: H.~D. III, A.~Singh (Eds.), Proceedings of
  the 37th International Conference on Machine Learning, Vol. 119 of
  Proceedings of Machine Learning Research, PMLR, Virtual, 2020, pp.
  10248--10259.

\bibitem{fort2020deep}
S.~Fort, H.~Hu, B.~Lakshminarayanan, Deep ensembles: A loss landscape
  perspective (2020).
\newblock \href {http://arxiv.org/abs/1912.02757} {\path{arXiv:1912.02757}}.

\bibitem{zaidi2020neural}
S.~Zaidi, A.~Zela, T.~Elsken, C.~Holmes, F.~Hutter, Y.~W. Teh, Neural ensemble
  search for performant and calibrated predictions (2020).
\newblock \href {http://arxiv.org/abs/2006.08573} {\path{arXiv:2006.08573}}.

\bibitem{10.5555/3295222.3295387}
B.~Lakshminarayanan, A.~Pritzel, C.~Blundell, Simple and scalable predictive
  uncertainty estimation using deep ensembles, in: Proceedings of the 31st
  International Conference on Neural Information Processing Systems, NIPS'17,
  Curran Associates Inc., Red Hook, NY, USA, 2017, p. 6405–6416.

\bibitem{ijcai2020-296}
M.~Wabartha, A.~Durand, V.~François-Lavet, J.~Pineau, Handling black swan
  events in deep learning with diversely extrapolated neural networks, in:
  C.~Bessiere (Ed.), Proceedings of the Twenty-Ninth International Joint
  Conference on Artificial Intelligence, {IJCAI-20}, International Joint
  Conferences on Artificial Intelligence Organization, 2020, pp. 2140--2147,
  main track.

\bibitem{UsingBMA}
A.~E. Raftery, T.~Gneiting, F.~Balabdaoui, M.~Polakowski, Using bayesian model
  averaging to calibrate forecast ensembles, Monthly Weather Review 133~(5) (01
  May. 2005) 1155 -- 1174.
\newblock \href {http://dx.doi.org/10.1175/MWR2906.1}
  {\path{doi:10.1175/MWR2906.1}}.

\bibitem{JSSv090i12}
A.~Jordan, F.~Krüger, S.~Lerch, Evaluating probabilistic forecasts with
  scoringrules, Journal of Statistical Software, Articles 90~(12) (2019) 1--37.
\newblock \href {http://dx.doi.org/10.18637/jss.v090.i12}
  {\path{doi:10.18637/jss.v090.i12}}.

\bibitem{doi:10.1146/annurev-statistics-062713-085831}
T.~Gneiting, M.~Katzfuss, Probabilistic forecasting, Annual Review of
  Statistics and Its Application 1~(1) (2014) 125--151.
\newblock \href {http://dx.doi.org/10.1146/annurev-statistics-062713-085831}
  {\path{doi:10.1146/annurev-statistics-062713-085831}}.

\bibitem{https://doi.org/10.1111/j.1467-9868.2007.00587.x}
T.~Gneiting, F.~Balabdaoui, A.~E. Raftery, Probabilistic forecasts, calibration
  and sharpness, Journal of the Royal Statistical Society: Series B
  (Statistical Methodology) 69~(2) (2007) 243--268.
\newblock \href
  {http://dx.doi.org/https://doi.org/10.1111/j.1467-9868.2007.00587.x}
  {\path{doi:https://doi.org/10.1111/j.1467-9868.2007.00587.x}}.

\bibitem{doi:10.1198/016214506000001437}
T.~Gneiting, A.~E. Raftery, Strictly proper scoring rules, prediction, and
  estimation, Journal of the American Statistical Association 102~(477) (2007)
  359--378.
\newblock \href {http://dx.doi.org/10.1198/016214506000001437}
  {\path{doi:10.1198/016214506000001437}}.

\bibitem{XU2019180}
L.~Xu, S.~Wang, R.~Tang,
  \href{http://www.sciencedirect.com/science/article/pii/S0306261919300224}{Probabilistic
  load forecasting for buildings considering weather forecasting uncertainty
  and uncertain peak load}, Applied Energy 237 (2019) 180 -- 195.
\newblock \href
  {http://dx.doi.org/https://doi.org/10.1016/j.apenergy.2019.01.022}
  {\path{doi:https://doi.org/10.1016/j.apenergy.2019.01.022}}.
\newline\urlprefix\url{http://www.sciencedirect.com/science/article/pii/S0306261919300224}

\bibitem{uk-smec}
Smartmeter energy consumption data in london households,
  \url{https://data.london.gov.uk/dataset/smartmeter-energy-use-data-in-london-households}.

\bibitem{adam}
D.~Kingma, J.~Ba, Adam: A method for stochastic optimization, International
  Conference on Learning Representations.

\bibitem{TF}
Tensorflow machine learning library, \url{https://www.tensorflow.org/}.

\bibitem{tfp}
Tensorflow probability library for probabilistic reasoning and statistical
  analysis, \url{https://www.tensorflow.org/probability}.

\end{thebibliography}

\end{document}